\documentclass[11pt]{article}
\usepackage{graphics}
\usepackage{hyperref}
\usepackage{amssymb}
\usepackage{color}

\title{\bf TASI 2013 lectures on \\
Higgs physics within and beyond the Standard Model\footnote{Comments are welcome and will inform future versions of these lecture notes.  I hereby release all the figures in these lectures into the public domain.  Share and enjoy!}}

\author{Heather E.\ Logan\thanks{\tt logan@physics.carleton.ca} \\
{\it Ottawa-Carleton Institute for Physics, Carleton University,}\\ 
{\it 1125 Colonel By Drive, Ottawa, Ontario K1S 5B6 Canada}}

\date{June 2014}

\begin{document}

\maketitle

\begin{abstract} 
\noindent
These lectures start with a detailed pedagogical introduction to electroweak symmetry breaking in the Standard Model, including gauge boson and fermion mass generation and the resulting predictions for Higgs boson interactions.  I then survey Higgs boson decays and production mechanisms at hadron and $e^+e^-$ colliders.  I finish with two case studies of Higgs physics beyond the Standard Model: two-Higgs-doublet models, which I use to illustrate the concept of minimal flavor violation, and models with isospin-triplet scalar(s), which I use to illustrate the concept of custodial symmetry.
\end{abstract}

\newpage
\tableofcontents
\newpage

\section{Introduction}

A Higgs boson was discovered at the CERN Large Hadron Collider (LHC) in July 2012~\cite{Hdiscovery}.  At the time of writing, its measured properties are consistent with those of the Standard Model (SM) Higgs boson, though deviations in its most important couplings at the level of ten(s of) percent are still possible, as is the existence of additional scalar particles making up an extended Higgs sector.  With this discovery, Higgs physics becomes an essential part of the education of any graduate student in high-energy physics.
These lectures are meant to be a pedagogical introduction to the Higgs mechanism\footnote{I do not make any attempt to relay the history of the development of the ideas that today go by the name of the Higgs mechanism.  A detailed account can be found in Chapter 11 of Sean Carroll's book, \emph{The Particle at the End of the Universe}~\cite{Carroll}.} in the SM and some of the resulting Higgs phenomenology.

These lectures are organized as follows.  In Sec.~\ref{sec:smHiggs} I give a detailed exposition of electroweak symmetry breaking in the SM, including the generation of the $W$ and $Z$ boson and fermion masses and the resulting predictions for Higgs boson interactions.  In Sec.~\ref{sec:pheno} I then survey the SM predictions for Higgs boson decays and production mechanisms at hadron and $e^+e^-$ colliders.  
In Secs.~\ref{sec:2hdm} and \ref{sec:triplets} I venture beyond the SM via two ``case studies'' of extended Higgs sectors.  Section~\ref{sec:2hdm} introduces the basic features of models containing two Higgs doublets, while Sec.~\ref{sec:triplets} introduces models containing isospin-triplet scalar field(s) in addition to the usual SM Higgs doublet.  Part of the purpose of these sections is to introduce the concepts of minimal flavor violation (in Sec.~\ref{sec:2hdm}) and custodial symmetry (in Sec.~\ref{sec:triplets}).  These two concepts are automatic features of the SM, and thus their important phenomenological consequences are best illustrated by comparison to models in which they are \emph{not} automatic.  
A brief philosophical outlook is given in Sec.~\ref{sec:summary}, followed by a collection of homework questions in Sec.~\ref{sec:homework}.

\section{The Higgs mechanism in the Standard Model}
\label{sec:smHiggs}

\subsection{Preliminaries: gauge sector}

Let's start with a review of the gauge and fermion parts of the SM Lagrangian.  The SM gauge structure is SU(3)$_c \times $SU(2)$_L \times $U(1)$_Y$, comprising respectively the strong interactions (subscript $c$ for color), weak isospin (subscript $L$ for the left-handed fermions it couples to), and hypercharge (subscript $Y$ for the hypercharge operator).  The gauge boson dynamics are encoded in the Lagrangian in terms of the field strength tensors:\footnote{I use the metric $g_{\mu\nu} = {\rm diag}(1, -1, -1, -1)$, so that $p^2 \equiv p_{\mu} p^{\mu} = m^2$ for an on-shell particle.}
\begin{equation}
	\mathcal{L}_{gauge} = - \frac{1}{4} G^a_{\mu\nu} G^{a \mu \nu} 
	- \frac{1}{4} W^a_{\mu\nu} W^{a \mu \nu}
	- \frac{1}{4} B_{\mu \nu} B^{\mu\nu},
\end{equation}
where repeated indices are always taken as summed.  Here the field strength tensors are given as follows.  For the U(1)$_Y$ interaction, the field strength tensor takes the same form as in electromagnetism,
\begin{equation}
	B_{\mu\nu} = \partial_{\mu} B_{\nu} - \partial_{\nu} B_{\mu}.
\end{equation}
For SU(3)$_c$, and non-abelian theories in general, the field strength tensor takes a more complicated form,
\begin{equation}
	G^a_{\mu\nu} = \partial_{\mu} G^a_{\nu} - \partial_{\nu} G^a_{\mu}
	+ g_s f^{abc} G^b_{\mu} G^c_{\nu},
\end{equation}
where $g_s$ is the strong interaction coupling strength, $a,b,c$ run from 1 to 8, and $f^{abc}$ are the (antisymmetric) \emph{structure constants} of SU(3), defined in terms of the group generators $t^a$ according to
\begin{equation}
	[t^a, t^b] = i f^{abc} t^c.
\end{equation}
For SU(2), $a,b,c$ run from 1 to 3 and $f^{abc} = \epsilon^{abc}$, the totally antisymmetric three-index tensor defined so that $\epsilon^{123} = 1$.  Therefore, the field strength tensor for SU(2)$_L$ can be written as
\begin{equation}
	W^a_{\mu\nu} = \partial_{\mu} W^a_{\nu} - \partial_{\nu} W^a_{\mu} 
	+ g \epsilon^{abc} W^b_{\mu} W^c_{\nu},
\end{equation}
where $g$ is the weak interaction coupling strength.

The gauge interactions of fermions or scalars are encoded in the \emph{covariant derivative},
\begin{equation}
	\mathcal{D}_{\mu} = \partial_{\mu} - i g^{\prime} B_{\mu} Y 
	- i g W^a_{\mu} T^a - i g_s G^a_{\mu} t^a,
	\label{eq:covariantderiv}
\end{equation}
where $g^{\prime}$ is the coupling strength of the hypercharge interaction, $Y$ is the hypercharge operator, and $T^a$ and $t^a$ are the SU(2) and SU(3) generators, respectively.  When acting upon a doublet representation of SU(2), $T^a$ is just $\sigma^a/2$ where $\sigma^a$ are the Pauli matrices,
\begin{equation}
	\sigma^1 = \left( \begin{array}{cc} 0 & 1 \\ 1 & 0 \end{array} \right), \qquad
	\sigma^2 = \left( \begin{array}{cc} 0 & -i \\ i & 0 \end{array} \right), \qquad
	\sigma^3 = \left( \begin{array}{cc} 1 & 0 \\ 0 & -1 \end{array} \right).
	\label{eq:paulimatrices}
\end{equation}

The corresponding gauge transformations can be written as follows (for the SU(2)$_L$ and SU(3)$_c$ gauge field transformations, we give only the infinitesimal form):  
\begin{eqnarray}
	{\rm U(1)}_Y: &\quad&
	\psi \to \exp [i \lambda_Y(x) Y] \psi, \qquad
	B_{\mu} \to B_{\mu} + \frac{1}{g^{\prime}} \partial_{\mu} \lambda_Y(x)
	\nonumber \\
	{\rm SU(2)}_L: &\quad&
	\psi \to \exp [i \lambda_L^a(x) T^a] \psi, \qquad
	W^a_{\mu} \to W^a_{\mu} + \frac{1}{g} \partial_{\mu} \lambda_L^a(x) + \epsilon^{abc} W^b_{\mu} \lambda_L^c(x) 
	\nonumber \\
	{\rm SU(3)}_c: &\quad&
	\psi \to \exp [i \lambda_c^a(x) t^a] \psi, \qquad
	G^a_{\mu} \to G^a_{\mu} + \frac{1}{g_s} \partial_{\mu} \lambda_c^a(x) + f^{abc} G^b_{\mu} \lambda_c^c(x).
	\label{eq:gaugetransformations}
\end{eqnarray}
A mass term for a gauge boson would take the form
\begin{equation}
	\mathcal{L} \supset \frac{1}{2} m_B^2 B_{\mu} B^{\mu}.
\end{equation}
This is \emph{not} gauge invariant and thus cannot be inserted by hand into the Lagrangian.  Therefore, (unbroken) gauge invariance implies that gauge bosons are all massless.  

\subsection{Preliminaries: fermion sector}

The SM contains three copies (generations) of a collection of \emph{chiral} fermion fields with different gauge transformation properties under SU(3)$_c \times $SU(2)$_L \times $U(1)$_Y$.  The content of a single generation is given in Table~\ref{tab:fermions}, along with their hypercharge assignments\footnote{A careful observer will notice that the electric charge of each field is given by $Q = T^3 + Y$.  We will derive this relationship in Sec.~\ref{sec:HiggsMech}.} (the value of the quantum number $Y$) and their SU(3)$_c$ (color) transformation properties.  The fields $Q_L$ and $L_L$ transform as doublets under SU(2)$_L$, while the remaining fields transform as singlets.  

\begin{table}
\begin{center}
\begin{tabular}{lccccc}
\hline\hline
 & $Q_L \equiv \left( \begin{array}{c} u_L \\ d_L \end{array} \right)$ & $u_R$ & $d_R$ & 
 $L_L \equiv \left( \begin{array}{c} \nu_L \\ e_L \end{array} \right)$ & $e_R$ \\
Hypercharge & $1/6$ & $2/3$ & $-1/3$ & $-1/2$ & $-1$ \\
Color & triplet & triplet & triplet & singlet & singlet \\
\hline \hline
 \end{tabular}
 \end{center}
 \caption{The chiral fermion content of a single generation of the Standard Model.}
 \label{tab:fermions}
 \end{table}

The left- and right-handed chiral fermion states are obtained from an unpolarized Dirac spinor using the \emph{projection operators}
\begin{equation}
	P_R = \frac{1}{2} (1 + \gamma^5), \qquad
	P_L = \frac{1}{2} (1 - \gamma^5),
\end{equation}
in such a way that
\begin{equation}
	P_R \psi \equiv \psi_R, \qquad
	P_L \psi \equiv \psi_L.
\end{equation}
Using the anticommutation relations $\{ \gamma^{\mu}, \gamma^5 \} = 0$ and the fact that $\gamma^5$ is Hermitian, we also have
\begin{equation}
	\bar \psi P_R = \psi^{\dagger} \gamma^0 P_R = \psi^{\dagger} P_L \gamma^0 = (P_L \psi)^{\dagger} \gamma^0 = \bar \psi_L,
\end{equation}
and similarly $\bar \psi P_L = \bar \psi_R$.  Finally, the projection operators obey $P_R + P_L = 1$ and $P_R^2 = P_R$, $P_L^2 = P_L$.

We can use this to rewrite the Dirac Lagrangian in terms of chiral fermion fields as follows.  We start with the Lagrangian for a generic fermion $\psi$ with mass $m$,
\begin{equation}
	\mathcal{L} = \bar \psi i \partial_{\mu} \gamma^{\mu} \psi - m \bar \psi \psi.
\end{equation}
The first term can be split into two terms involving left- and right-handed chiral fermion fields by inserting a factor of $1 = (P_L^2 + P_R^2)$ before the $\psi$ and using the anticommutation relation to pull one factor of the projection operator through the $\gamma^{\mu}$ in each term:
\begin{equation}
	\bar \psi i \partial_{\mu} \gamma^{\mu} \psi 
	= \bar \psi P_R i \partial_{\mu} \gamma^{\mu} P_L \psi
	+ \bar \psi P_L i \partial_{\mu} \gamma^{\mu} P_R \psi
	= \bar \psi_L i \partial_{\mu} \gamma^{\mu} \psi_L 
	+ \bar \psi_R i \partial_{\mu} \gamma^{\mu} \psi_R.
\end{equation}
The kinetic term separates neatly into one term involving only $\psi_L$ and one involving only $\psi_R$.  We can then incorporate the gauge transformation properties by promoting the derivative $\partial_{\mu}$ to a covariant derivative $\mathcal{D}_{\mu}$ and these two terms will be gauge invariant for any of the fermion fields given in Table~\ref{tab:fermions}.

Now let's consider the mass term.  Using the same tricks, we have,
\begin{equation}
	-m \bar \psi \psi = -m \bar \psi P_L^2 \psi - m \bar \psi P_R^2 \psi
	= -m \bar \psi_R \psi_L - m \bar \psi_L \psi_R.
	\label{eq:LR+RL}
\end{equation}
(Note that the second term is just the Hermitian conjugate of the first term.)  The mass terms each involve fermions of \emph{both} chiralities.  Because the left-handed and right-handed fermions of the SM carry different SU(2)$_L \times$U(1)$_Y$ gauge charges, such mass terms are \emph{not} gauge invariant and thus cannot be inserted by hand into the Lagrangian.  Therefore, given the gauge charges of the SM fermions, (unbroken) gauge invariance implies that all the SM fermions are massless.\footnote{Some models beyond the SM contain left- and right-handed chiral fermions that carry the same SU(2)$_L \times$U(1)$_Y$ gauge charges, and can thus form a massive Dirac fermion without any reference to electroweak symmetry breaking.  Such fermions are called \emph{vectorlike fermions}, because of their pure vector (as opposed to axial-vector) couplings to the $Z$ boson.}

\subsection{The SM Higgs mechanism}
\label{sec:HiggsMech}

We have established that the theoretical explanation of the experimentally-observed nonzero masses of the $W$ and $Z$ bosons and the SM fermions requires a new ingredient.  Such an explanation is achieved by introducing a single SU(2)$_L$-doublet scalar field, which causes spontaneous breaking of the SU(2)$_L \times$U(1)$_Y$ gauge symmetry via the \emph{Higgs mechanism}.

We add to the SM a field $\Phi$, an SU(2)$_L$-doublet of complex scalar fields that can be written as
\begin{equation}
	\Phi = \left( \begin{array}{c} \phi^+  \\ \phi^0 \end{array} \right)
	= \frac{1}{\sqrt{2}} \left( \begin{array}{c} \phi_1 + i \phi_2 \\ \phi_3 + i \phi_4 \end{array} \right),
	\label{eq:phi4component}
\end{equation}
where $\phi_1, \phi_2, \phi_3, \phi_4$ are properly normalized real scalar fields.  We assign $\Phi$ a hypercharge $Y = 1/2$ and make it a color singlet.  The new terms in the Lagrangian involving $\Phi$ are given by
\begin{equation}
	\mathcal{L}_{\Phi} = ( \mathcal{D}_{\mu} \Phi)^{\dagger} (\mathcal{D}^{\mu} \Phi) 
	- V(\Phi) + \mathcal{L}_{\rm Yukawa},
\end{equation}
where the first term contains the kinetic and gauge-interaction terms via the covariant derivative, the second term is a potential energy function involving $\Phi$, and the third term contains Yukawa couplings of the scalar field to pairs of fermions.  We will treat each term in turn, starting with the potential energy function.

\begin{figure}
\begin{center}
\resizebox{0.5\textwidth}{!}{\includegraphics{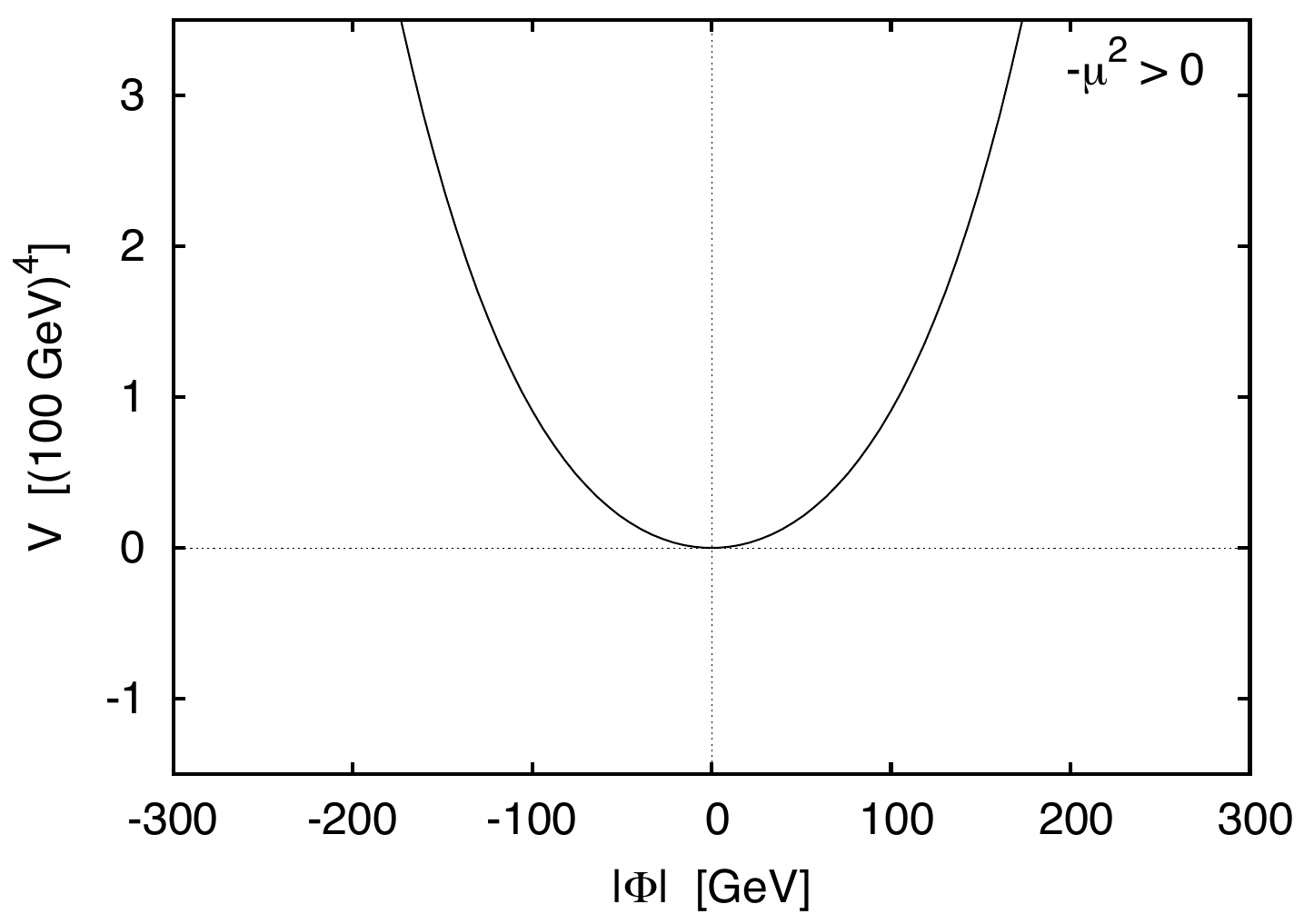}}\resizebox{0.5\textwidth}{!}{\includegraphics{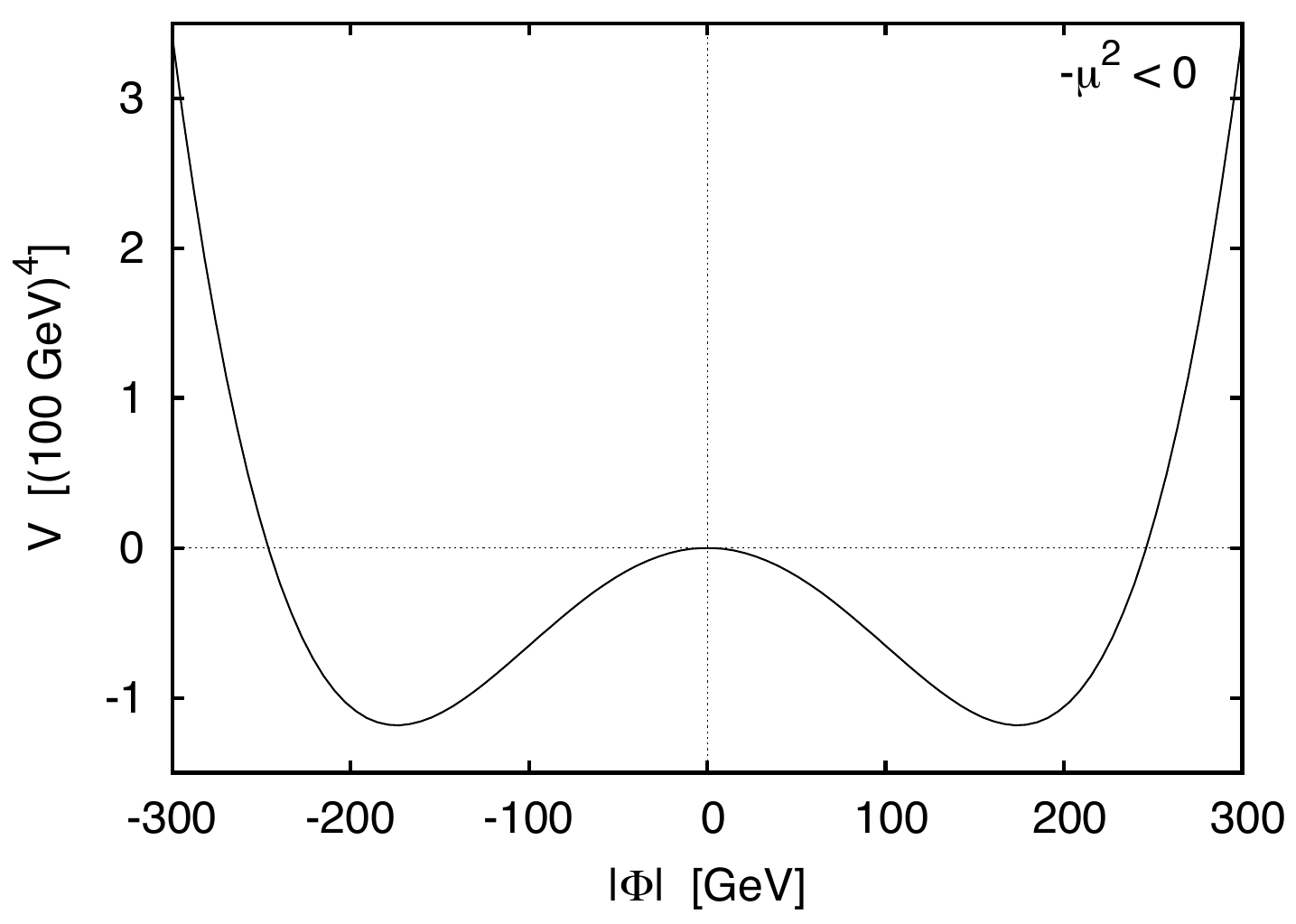}}
\end{center}
\caption{Plots of $V(\Phi) = -\mu^2 \Phi^{\dagger} \Phi + \lambda (\Phi^{\dagger} \Phi)^2$ as a function of $|\Phi| \equiv \sqrt{\Phi^{\dagger}\Phi}$ for the cases $-\mu^2 > 0$ (left) and $-\mu^2 < 0$ (right).  For the SM parameters I used $|{-}\mu^2| \simeq (88.4~{\rm GeV})^2$ and $\lambda \simeq 0.129$, obtained from the measured values $m_h \simeq 125$~GeV and $v \simeq 246$~GeV.  In the case that $-\mu^2 < 0$ (right), the minimum of the potential is at $|\Phi| = v/\sqrt{2} = (246/\sqrt{2})$~GeV.}
\label{fig:V}
\end{figure}

The most general gauge invariant potential energy function, or \emph{scalar potential}, involving $\Phi$ is given by
\begin{equation}
	V(\Phi) = - \mu^2 \Phi^{\dagger} \Phi + \lambda (\Phi^{\dagger} \Phi)^2.
	\label{eq:V}
\end{equation}
Consider the possible signs of the coefficients of the two terms in $V$:
\begin{itemize}
\item If $\lambda$ is negative, then $V$ is unbounded from below and there is no stable vacuum state.
\item When $-\mu^2$ and $\lambda$ are both positive, the potential energy function has a minimum at $|\Phi| \equiv \sqrt{\Phi^{\dagger} \Phi} = 0$ (left panel of Fig.~\ref{fig:V}).  In this case the electroweak symmetry is unbroken in the vacuum, because a gauge transformation acting on the vacuum state $\Phi = 0$ does not change the vacuum state.
\item When $-\mu^2$ is negative and $\lambda$ is positive, the potential energy function has a minimum away from $|\Phi| = 0$ (right panel of Fig.~\ref{fig:V}).  In this case the vacuum, or minimum energy state, is not invariant under SU(2)$_L \times$U(1)$_Y$ transformations: the gauge symmetry is \emph{spontaneously broken} in the vacuum.
\end{itemize}

Let's take a closer look at the symmetry-breaking case.  The Higgs field $\Phi$ is a complex scalar field with two isospin components; we can thus write it in terms of four real scalar degrees of freedom as in Eq.~(\ref{eq:phi4component}),
where the $1/\sqrt{2}$ normalization ensures that the kinetic energy terms for the real scalars will have the correct normalization, $\mathcal{L} \supset \frac{1}{2} \partial_{\mu} \phi_i \partial^{\mu} \phi_i$.  Then
\begin{equation}
	\Phi^{\dagger} \Phi = \frac{1}{2} \left( \phi_1^2 + \phi_2^2 + \phi_3^2 + \phi_4^2 \right),
\end{equation}
which can be thought of as the square of the length of a four-component vector.  Minimizing the potential in Eq.~(\ref{eq:V}) fixes the length of this vector to satisfy
\begin{equation}
	\Phi^{\dagger} \Phi = \frac{\mu^2}{2 \lambda},
\end{equation}
which is a positive quantity when $-\mu^2$ is negative.  This picks out a spherical surface in four dimensions upon which the potential is minimized.\footnote{For the topologically inclined, the vacuum manifold is $S^3$.}

In this language, SU(2)$_L \times$U(1)$_Y$ gauge transformations correspond to rotations in this four-dimensional space.\footnote{Note that there are four independent SU(2)$_L \times$U(1)$_Y$ gauge transformations in Eq.~(\ref{eq:gaugetransformations}) but only three independent rotation directions for a vector in a four-dimensional space.  In fact, there is always one combination of the SU(2)$_L$ and U(1)$_Y$ transformations that leaves the vacuum state invariant.  This particular combination of gauge transformations will remain unbroken by the Higgs field and corresponds to the gauge transformation of electromagnetism.}  Under such rotations $V$ is invariant---the value of the potential depends only on the distance from the origin---but a particular vacuum state (a particular vector of length $\sqrt{\mu^2/2 \lambda}$) transforms nontrivially: it is rotated into a new vector of the same length but pointing in a different direction.

We also acquire a physical picture for excitations around such a vacuum state.  Excitations in any of the three rotational directions cost zero energy, because the potential is flat in those directions.  These correspond to massless modes or Goldstone modes.  An excitation in the radial direction, on the other hand, feels an approximate harmonic oscillator potential about the minimum and gives rise to a massive particle.

Let's see how this works explicitly.  The potential is given in terms of the four real scalars by
\begin{equation}
	V = - \frac{\mu^2}{2} \left( \phi_1^2 + \phi_2^2 + \phi_3^2 + \phi_4^2 \right) 
	+ \frac{\lambda}{4} \left( \phi_1^2 + \phi_2^2 + \phi_3^2 + \phi_4^2 \right)^2.
\end{equation}
We are free to choose the basis of states $\phi_1, \cdots, \phi_4$ to be oriented however we like relative to the local vacuum value; let's choose the vacuum expectation values (``vevs'') of the four fields to be
\begin{equation}
	\langle \phi_3 \rangle \equiv v = \sqrt{\frac{\mu^2}{\lambda}}, \qquad \qquad
	\langle \phi_1 \rangle = \langle \phi_2 \rangle = \langle \phi_4 \rangle = 0.
\end{equation}
We can also define a new real scalar field $h$ with zero vacuum value, $\langle h \rangle = 0$, according to 
\begin{equation}
	\phi_3 = h + v.
\end{equation}
Then our field becomes
\begin{equation}
	\Phi = \frac{1}{\sqrt{2}} \left( \begin{array}{c} \phi_1 + i \phi_2 \\
		v + h + i \phi_4 \end{array} \right),
	\label{eq:Phi4component}
\end{equation}
and the potential becomes
\begin{equation}
	V = - \frac{\mu^2}{2} \left( \phi_1^2 + \phi_2^2 + (h + v)^2 + \phi_4^2 \right) 
	+ \frac{\lambda}{4} \left( \phi_1^2 + \phi_2^2 + (h + v)^2 + \phi_4^2 \right)^2.
\end{equation}
In particular, we have expressed the potential entirely in terms of constants and fields with zero vacuum value.  This lets us treat the fields in terms of small excitations as usual in quantum field theory.  Multiplying out the terms in $V$ and using $\mu^2 = \lambda v^2$ to eliminate $\mu^2$, we find
\begin{equation}
	V = {\rm constant} + 0 \cdot \phi_1^2 + 0 \cdot \phi_2^2 + \lambda v^2 h^2 + 0 \cdot \phi_4^2
	+ {\rm cubic} + {\rm quartic}.
\end{equation}
These quadratic terms are the mass terms for the real scalars.  We see that $\phi_1$, $\phi_2$, and $\phi_4$ are massless in accordance with our intuitive picture above, while $h$ has a mass $m_h = \sqrt{2 \lambda v^2}$.\footnote{Recall that for a real scalar $\phi$ with mass $m$, $V \supset \frac{1}{2} m^2 \phi^2$.}

To learn more about the nature of the massless modes, we can rewrite $\Phi$ in another convenient form,
\begin{equation}
	\Phi = \frac{1}{\sqrt{2}} \exp \left( \frac{i \xi^a \sigma^a}{v} \right) \left( \begin{array}{c} 0 \\
		v + h \end{array} \right).
\end{equation}
Here $h$ and $\xi^a$ are fields, $\sigma^a$ are the Pauli matrices as in Eq.~(\ref{eq:paulimatrices}), and $a$ is summed over $1,2,3$.  This expression is equivalent to Eq.~(\ref{eq:Phi4component}) up to linear order in the fields, i.e., for infinitesimal fluctuations about the vacuum.\footnote{To linear order, $\xi^1 = \phi_2$, $\xi^2 = \phi_1$, and $\xi^3 = - \phi_4$.}

Now consider the gauge transformations of $\Phi$:
\begin{eqnarray}
	{\rm U(1)}_Y: &\quad& \Phi \to \exp \left( i \lambda_Y(x) \cdot \frac{1}{2} \right) \Phi, \nonumber \\
	{\rm SU(2)}_L: &\quad& \Phi \to \exp \left( i \lambda_L^a(x) \frac{\sigma^a}{2} \right) \Phi.
\end{eqnarray}
If we choose $\lambda_L^a(x) = - 2 \xi^a/v$ at each point in spacetime, we arrive at a gauge in which
\begin{equation}
	\Phi = \frac{1}{\sqrt{2}} \left( \begin{array}{c} 0 \\ v + h \end{array} \right),
\end{equation}
i.e., we have \emph{gauged away} the fields $\xi^a$, or equivalently $\phi_1, \phi_2, \phi_4$.\footnote{Note that we could have gauged away $\xi^3$ by doing an appropriate combination of SU(2)$_L$ and U(1)$_Y$ gauge transformations.}  These fields have been entirely removed from the Lagrangian by means of a gauge transformation!\footnote{This removal of the Goldstone modes by means of a gauge transformation is sometimes described as the Goldstones being ``eaten'' by the corresponding gauge bosons.}  This means that it must be possible to interpret the theory in a way in which these fields are absent (but with the gauge fixed): they are not physical degrees of freedom.  This gauge choice is known as unitary or unitarity gauge.  The massive field $h$ remains present and always shows up in the combination $(v + h)$.

\subsection{Gauge boson masses and couplings to the Higgs boson}

We now examine the gauge-kinetic term,
\begin{equation}
	\mathcal{L} \supset \left( \mathcal{D}_{\mu} \Phi \right)^{\dagger} \left( \mathcal{D}^{\mu} \Phi \right).
\end{equation}
When acting on $\Phi$, the covariant derivative reads
\begin{equation}
	\mathcal{D}_{\mu} = \partial_{\mu} - i \frac{g^{\prime}}{2} B_{\mu} - i \frac{g}{2} W_{\mu}^a \sigma^a.
\end{equation}
Applying this to $\Phi$ in the unitarity gauge we find
\begin{equation}
	\mathcal{D}_{\mu} \Phi = \frac{1}{\sqrt{2}} \left( \begin{array}{c}
	- \frac{i}{2} g (W^1_{\mu} - i W^2_{\mu}) (v + h) \\
	\partial_{\mu} h + \frac{i}{2} ( g W^3_{\mu} - g^{\prime} B_{\mu} ) (v + h)
	\end{array} \right).
\end{equation}
Dotting this into its Hermitian conjugate gives,
\begin{equation}
	\left( \mathcal{D}_{\mu} \Phi \right)^{\dagger} \left( \mathcal{D}^{\mu} \Phi \right)
	= \frac{1}{2} ( \partial_{\mu} h ) (\partial^{\mu} h)
	+ \frac{1}{8} g^2 (v + h)^2 (W^1_{\mu} - i W^2_{\mu}) (W^{1 \mu} + i W^{2 \mu})
	+ \frac{1}{8} (v + h)^2 \left( -g^{\prime} B_{\mu} + g W^3_{\mu} \right)^2.
	\label{eq:gaugekinetic}
\end{equation}

\begin{figure}
\begin{center}
\includegraphics{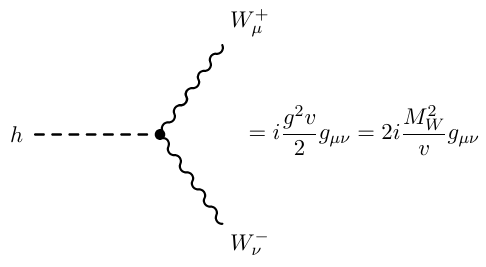} \hspace*{\fill}
\includegraphics{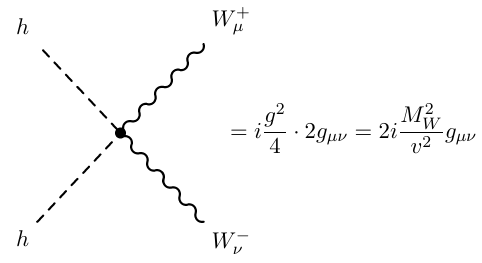}
\end{center}
\caption{Feynman rules for the $hWW$ and $hhWW$ vertices, as derived from the Lagrangian in Eq.~(\ref{eq:Whcoups}).  The extra factor of $2$ in the first expression for the $hhWW$ coupling is a symmetry factor accounting for the two identical Higgs bosons.  See also Eq.~(\ref{eq:hww}).}
\label{fig:hww}
\end{figure}

Let us consider the three terms in turn.  The first is the properly normalized kinetic term for the real scalar field $h$ (the Higgs boson).  For the second term, we note that the combinations $W^1 \pm i W^2$ correspond to the charged $W$ bosons:\footnote{Which combination corresponds to $W^+$ and which to $W^-$?  This can be checked by noting that 
\begin{equation}
	W^1_{\mu} \sigma^1 + W^2_{\mu} \sigma^2 
	= \frac{1}{2} (W^1_{\mu} - i W^2_{\mu}) (\sigma^1 + i \sigma^2) 
	+ \frac{1}{2} (W^1_{\mu} + i W^2_{\mu}) (\sigma^1 - i \sigma^2) 
	= \sqrt{2} \frac{W^1_{\mu} - i W^2_{\mu}}{\sqrt{2}} \sigma^+
	+ \sqrt{2} \frac{W^1_{\mu} + i W^2_{\mu}}{\sqrt{2}} \sigma^-,
\end{equation}
where 
\begin{equation}
	(\sigma^1 + i \sigma^2) = 2 \sigma^+ = 2 \left( \begin{array}{cc} 0 & 1 \\ 0 & 0 \end{array} \right),
	\qquad
	(\sigma^1 - i \sigma^2) = 2 \sigma^- = 2 \left( \begin{array}{cc} 0 & 0 \\ 1 & 0 \end{array} \right).
\end{equation}
When the covariant derivative acts on the left-handed fermion doublets we get terms of the following form, from which we can identify $W^+$ and $W^-$ using charge conservation:
\begin{eqnarray}
	\frac{W^1_{\mu} - i W^2_{\mu}}{\sqrt{2}} \left( \bar u \ \bar d \, \right) \sigma^+ 
		\gamma^{\mu} P_L \left( \begin{array}{c} u \\ d \end{array} \right) 
	&=& \frac{W^1_{\mu} - i W^2_{\mu}}{\sqrt{2}} \bar u \gamma^{\mu} P_L d  
	\qquad \Rightarrow \qquad
	\frac{W^1_{\mu} - i W^2_{\mu}}{\sqrt{2}} = W^+_{\mu},
	\nonumber \\
	\frac{W^1_{\mu} + i W^2_{\mu}}{\sqrt{2}} \left( \bar u \ \bar d \, \right) \sigma^-
		\gamma^{\mu} P_L \left( \begin{array}{c} u \\ d \end{array} \right) 
	&=& \frac{W^1_{\mu} + i W^2_{\mu}}{\sqrt{2}} \bar d \gamma^{\mu} P_L u  
	\qquad \Rightarrow \qquad
	\frac{W^1_{\mu} + i W^2_{\mu}}{\sqrt{2}} = W^-_{\mu}.
\end{eqnarray}}
\begin{equation}
	W^+_{\mu} = \frac{W^1_{\mu} - i W^2_{\mu}}{\sqrt{2}}, \qquad
	W^-_{\mu} = \frac{W^1_{\mu} + i W^2_{\mu}}{\sqrt{2}}.
\end{equation}
The second term in Eq.~(\ref{eq:gaugekinetic}) becomes
\begin{eqnarray}
	\mathcal{L} &\supset& \frac{1}{8} g^2 (v + h)^2 (W^1_{\mu} - i W^2_{\mu}) (W^{1 \mu} + i W^{2 \mu}) \nonumber \\
	&=& \frac{1}{4} g^2 (v + h)^2 W^+_{\mu} W^{- \mu} \nonumber \\
	&=& \frac{g^2 v^2}{4} W^+_{\mu} W^{- \mu}
	+ \frac{g^2 v}{2} h W^+_{\mu} W^{- \mu}
	+ \frac{g^2}{4} h h W^+_{\mu} W^{- \mu}.
	\label{eq:Whcoups}
\end{eqnarray}
The first term here is a mass term for the $W$ boson, with
\begin{equation}
	M_W^2 = \frac{g^2 v^2}{4}.
\end{equation}
The Higgs vacuum expectation value (vev) has given the $W$ boson a mass!  Because $M_W$ and $g$ have been directly measured, we can determine $v \simeq 246$~GeV.\footnote{This value of $v$ actually comes from the Fermi constant, $G_F = 1/\sqrt{2} v^2$.}
The second and third terms in Eq.~(\ref{eq:Whcoups}) give interactions of one or two Higgs bosons with $W^+W^-$.  The corresponding Feynman rules (see Fig.~\ref{fig:hww}) are
\begin{eqnarray}
	h W^+_{\mu} W^-_{\nu}: &\quad& i \frac{g^2 v}{2} g_{\mu\nu} 
	= i g M_W g_{\mu \nu}
	= 2 i \frac{M_W^2}{v} g_{\mu\nu},
	\nonumber \\
	h h W^+_{\mu} W^-_{\nu}: &\quad& i \frac{g^2}{4} \times 2! \, g_{\mu\nu} 
	= 2 i \frac{M_W^2}{v^2} g_{\mu\nu},
	\label{eq:hww}
\end{eqnarray}
where the $2!$ in the second expression is a combinatorical factor from the two identical Higgs bosons in the Lagrangian term.
Note that the $W$ mass, the $hWW$ coupling, and the $hhWW$ coupling all come from the same term in the Lagrangian and are generated by expanding out the factor $(v + h)^2$.  Thus the $hWW$ and $hhWW$ couplings are uniquely predicted in the SM once the $W$ mass and $v$ are known.

We now consider the third term of Eq.~(\ref{eq:gaugekinetic}).  We first write the linear combination of $W^3_{\mu}$ and $B_{\mu}$ that appears in this term as a properly normalized real field:
\begin{eqnarray}
	\left( g W^3_{\mu} - g^{\prime} B_{\mu} \right) 
	&=& \sqrt{g^2 + g^{\prime 2}} \left( \frac{g}{\sqrt{g^2 + g^{\prime 2}}} W^3_{\mu}
	- \frac{g^{\prime}}{\sqrt{g^2 + g^{\prime 2}}} B_{\mu} \right) \nonumber \\
	&\equiv& \sqrt{g^2 + g^{\prime 2}} \left( c_W W^3_{\mu} - s_W B_{\mu} \right) \nonumber \\
	&\equiv& \sqrt{g^2 + g^{\prime 2}} \, Z_{\mu},
\end{eqnarray}
where we have defined $s_W = \sin \theta_W$, $c_W = \cos\theta_W$, where $\theta_W$ is the weak mixing angle or Weinberg angle.  We have also defined the field combination $Z_{\mu}$, which will receive a mass from the Higgs vev and be identified as the $Z$ boson.

We note that the orthogonal state,
\begin{equation}
	\left( s_W W^3_{\mu} + c_W B_{\mu} \right) \equiv A_{\mu},
\end{equation}
does not couple to the Higgs field and thus does not acquire a mass through the Higgs mechanism.  This state will be identified as the photon.\footnote{The choice of basis of the Higgs field, i.e., in which component we put the vev, does not affect this conclusion.  There will always remain one massless gauge boson, corresponding to the combination of SU(2)$_L$ and U(1)$_Y$ gauge transformations that leaves our chosen vacuum state invariant.  This combination will not couple to $(v + h)^2$, will not acquire a mass, and will thus be identified with the known massless electroweak gauge boson, the photon.  Since electric charge is defined in terms of the couplings of the photon, the SM Higgs vev and physical Higgs boson will always be what we call electrically neutral.}

\begin{figure}
\begin{center}
\includegraphics{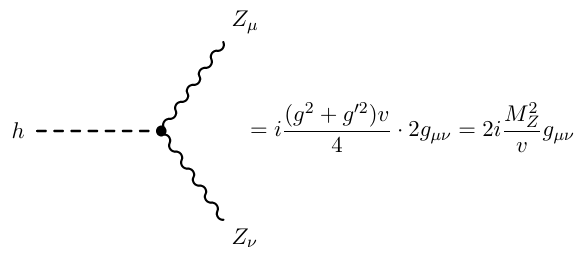} 
\includegraphics{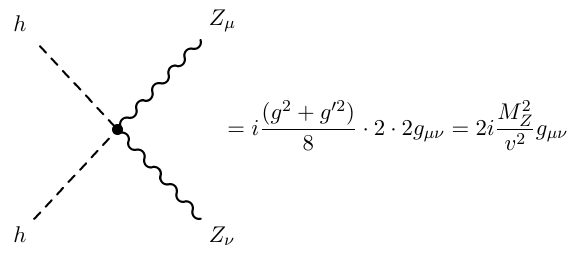}
\end{center}
\caption{Feynman rules for the $hZZ$ and $hhZZ$ vertices, as derived from the Lagrangian in Eq.~(\ref{eq:Zhcoups}).  The extra factor of $2$ in the first expression for the $hZZ$ coupling is a symmetry factor accounting for the two identical $Z$ bosons.  The $hhZZ$ coupling contains two extra factors of $2$ which are the symmetry factors accounting respectively for the two identical Higgs bosons and two identical $Z$ bosons.  See also Eq.~(\ref{eq:hzz}).}
\label{fig:hzz}
\end{figure}

The third term in Eq.~(\ref{eq:gaugekinetic}) becomes
\begin{eqnarray}
	\mathcal{L} &\supset& \frac{1}{8} (v + h)^2 \left( -g^{\prime} B_{\mu} + g W^3_{\mu} \right)^2
	\nonumber \\
	&=& \frac{1}{8} (g^2 + g^{\prime 2}) (v + h)^2 Z_{\mu} Z^{\mu} \nonumber \\
	&=& \frac{(g^2 + g^{\prime 2}) v^2}{8} Z_{\mu} Z^{\mu}
	+ \frac{(g^2 + g^{\prime 2}) v}{4} h Z_{\mu} Z^{\mu}
	+ \frac{(g^2 + g^{\prime 2})}{8} h h Z_{\mu} Z^{\mu}.
	\label{eq:Zhcoups}
\end{eqnarray}
The first term here is a mass term for the $Z$ boson,\footnote{Remember that the mass term for a real vector field takes the form $\mathcal{L} \supset \frac{1}{2} M_Z^2 Z_{\mu} Z^{\mu}$.}
\begin{equation}
	M_Z^2 = \frac{(g^2 + g^{\prime 2}) v^2}{4}.
\end{equation}
The second and third terms in Eq.~(\ref{eq:Zhcoups}) give interactions of one or two Higgs bosons with $ZZ$.  The corresponding Feynman rules (see Fig.~\ref{fig:hzz}) are
\begin{eqnarray}
	h Z_{\mu} Z_{\nu}: &\quad& i \frac{(g^2 + g^{\prime 2}) v}{4} \times 2! \, g_{\mu\nu}
	= i \sqrt{g^2 + g^{\prime 2}} M_Z g_{\mu\nu}
	= 2 i \frac{M_Z^2}{v} g_{\mu\nu}, \nonumber \\
	hh Z_{\mu} Z_{\nu}: &\quad& i \frac{(g^2 + g^{\prime 2})}{8} \times 2! \times 2! \, g_{\mu\nu}
	= 2 i \frac{M_Z^2}{v^2} g_{\mu\nu},
	\label{eq:hzz}
\end{eqnarray}
where each coupling contains a $2!$ from the two identical $Z$ bosons, and the second expression contains an extra $2!$ from the two identical Higgs bosons in the Lagrangian term.
As before, the $Z$ mass, the $hZZ$ coupling, and the $hhZZ$ coupling all come from the same term in the Lagrangian and are generated by expanding out the factor $(v + h)^2$.  Thus the $hZZ$ and $hhZZ$ couplings are uniquely predicted in the SM once the $Z$ mass and $v$ are known.

We can now rewrite the covariant derivative in terms of our new basis of electroweak gauge bosons, $W^+$, $W^-$, $Z$, and $A$.  Starting from Eq.~(\ref{eq:covariantderiv}), we make the following substitutions:
\begin{eqnarray}
	B_{\mu} &=& c_W A_{\mu} - s_W Z_{\mu}, \nonumber \\
	W^3_{\mu} &=& s_W A_{\mu} + c_W Z_{\mu}, \nonumber \\
	W^1 T^1 + W^2 T^2 &=& \frac{1}{\sqrt{2}} ( W^+ T^+ + W^- T^- ),
\end{eqnarray}
where $T^{\pm}$ are the raising and lowering operators of SU(2)$_L$, with $T^{\pm} = \sigma^{\pm}$ in the doublet representation.  This yields,
\begin{equation}
	\mathcal{D}_{\mu} = \partial_{\mu} - i g_s G^a_{\mu} t^a
	- i \frac{g}{\sqrt{2}} \left( W^+_{\mu} T^+ + W^-_{\mu} T^- \right)
	- i Z_{\mu} \left( g c_W T^3 - g^{\prime} s_W Y \right)
	- i A_{\mu} \left( g s_W T^3 + g^{\prime} c_W Y \right).
\end{equation}

We first examine the photon coupling.  Using the definitions $s_W = g^{\prime}/\sqrt{g^2 + g^{\prime 2}}$, $c_W = g/\sqrt{g^2 + g^{\prime 2}}$, we can simplify the coefficient 
\begin{equation}
	\left( g s_W T^3 + g^{\prime} c_W Y \right) 
	= \frac{g g^{\prime}}{\sqrt{g^2 + g^{\prime 2}}} \left( T^3 + Y \right) 
	\equiv e Q,
\end{equation}
where $e$ is the electromagnetic coupling and $Q$ is the electric charge operator.  By convention, we identify
\begin{equation}
	e = \frac{g g^{\prime}}{\sqrt{g^2 + g^{\prime 2}}} = g s_W = g^{\prime} c_W,
	\qquad \qquad
	Q = T^3 + Y.
\end{equation}
The photon coupling then takes the familiar form $\mathcal{D}_{\mu} \supset -i e A_{\mu} Q$.

Now let's examine the $Z$ boson coupling.  We can use $Y = Q - T^3$ to write
\begin{equation}
	\left( g c_W T^3 - g^{\prime} s_W Y \right) 
	= \frac{g^2 + g^{\prime 2}}{\sqrt{g^2 + g^{\prime 2}}} T^3 - \frac{g^{\prime 2}}{\sqrt{g^2 + g^{\prime 2}}} Q
	= \sqrt{g^2 + g^{\prime 2}} \left( T^3 - s_W^2 Q \right).
\end{equation}

Putting it all together, we obtain the covariant derivative in the gauge boson mass basis,
\begin{equation}
	\mathcal{D}_{\mu} = \partial_{\mu} - i g_s G^a_{\mu} t^a
	- i \frac{g}{\sqrt{2}} \left( W^+_{\mu} T^+ + W^-_{\mu} T^- \right)
	- i \frac{e}{s_Wc_W} Z_{\mu} \left( T^3 - s_W^2 Q \right)
	- i e A_{\mu} Q,
\end{equation}
where we note that $g = e/s_W$ and $e/s_W c_W = g/c_W = \sqrt{g^2 + g^{\prime 2}}$.  From this expression we can derive the familiar electroweak fermion-antifermion-gauge boson Feynman rules using the fermion gauge-kinetic terms,
\begin{equation}
	\mathcal{L} \supset \bar \psi_L i \mathcal{D}_{\mu} \gamma^{\mu} \psi_L
	+ \bar \psi_R i \mathcal{D}_{\mu} \gamma^{\mu} \psi_R.
\end{equation}

\subsection{Fermion masses, the CKM matrix, and couplings to the Higgs boson}

Now let's look at the couplings of the Higgs doublet $\Phi$ to fermions.  We'll start with the leptons and neglect neutrino masses\footnote{I'll make some comments on neutrino masses later in this subsection.} for simplicity.

\subsubsection{Lepton masses}

The construction of the Lagrangian terms that describe the Higgs couplings to fermions is pretty straightforward.  Lorentz invariance (conservation of spin) requires that fermion spinors appear in pairs, $\bar \psi \psi$.  Because the fermion field has mass dimension $3/2$, $\bar \psi \psi$ has mass dimension 3; combining this with a single Higgs doublet (with mass dimension 1) already yields mass dimension 4.  Thus we can construct renormalizable Higgs-fermion couplings involving only one each of $\bar \psi$, $\psi$, and $\Phi$.  Furthermore, $\Phi$ is an SU(2)$_L$ doublet; for our Lagrangian term to be gauge invariant, we must couple it to one SU(2)$_L$ doublet fermion field (e.g., $L_L = (\nu_L, e_L)^T$, see Table~\ref{tab:fermions}) and one SU(2)$_L$ singlet (e.g., $e_R$).

Following this logic, the most general gauge-invariant renormalizable Lagrangian terms involving the Higgs doublet and leptons are, for a single generation,\footnote{You can add up the hypercharges of the fields in these Lagrangian terms, remembering that a Hermitian-conjugated field carries minus the hypercharge of the original field, and see that the net hypercharge of each term is zero, i.e., that these terms are also gauge invariant under U(1)$_Y$.  The same is true for the up- and down-type quark Yukawa terms that we will write down below.  Aren't we lucky that the hypercharges of the left-handed fermions, right-handed fermions, and Higgs doublet work out just right to allow for the generation of fermion masses via electroweak symmetry breaking!  Why this works out so nicely is a mystery in the SM, possibly to be explained by grand unification of the gauge interactions.}
\begin{equation}
	\mathcal{L}_{\rm Yukawa} \supset - \left[ y_e \bar e_R \Phi^{\dagger} L_L 
	+ y_e^* \bar L_L \Phi e_R \right],
\end{equation}
where the second term is the Hermitian conjugate of the first and $y_e$ is a dimensionless constant.  The coupling $y_e$ is complex in general, but its phase can be absorbed into a physically-undetectable rephasing of the right-handed electron field $e_R$; therefore we'll treat it as real in what follows.

\begin{figure}
\begin{center}
\includegraphics{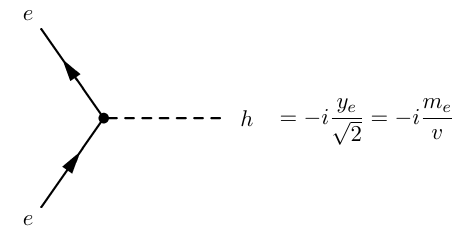}
\end{center}
\caption{Feynman rule for the $h \bar ee$ vertex, as derived from the Lagrangian in Eq.~(\ref{eq:hee}).  See also Eq.~(\ref{eq:heefr}).}
\label{fig:hee}
\end{figure}

In unitarity gauge,
\begin{equation}
	\Phi = \left( \begin{array}{c} 0 \\ (v + h)/\sqrt{2} \end{array} \right),
\end{equation}
and
\begin{equation}
	\Phi^{\dagger} L_L = \left( 0, \frac{v + h}{\sqrt{2}} \right) 
	\left( \begin{array}{c} \nu_e \\ e \end{array} \right)_L 
	= \frac{v + h}{\sqrt{2}} e_L,
\end{equation}
so [using Eq.~(\ref{eq:LR+RL}) in the second step]
\begin{eqnarray}
	\mathcal{L}_{\rm Yukawa} &\supset& - y_e \frac{1}{\sqrt{2}} 
	\left[ (v + h) \bar e_R e_L + (v + h) \bar e_L e_R \right] \nonumber \\
	&=& - \frac{y_e}{\sqrt{2}} (v + h) \bar e e \nonumber \\
	&=& - \left( \frac{y_e v}{\sqrt{2}} \right) \bar e e - \frac{y_e}{\sqrt{2}} h \bar e e.
	\label{eq:hee}
\end{eqnarray}
The first term in the last line is a mass term for the electron,
\begin{equation}
	m_e = \frac{y_e v}{\sqrt{2}}.
\end{equation}
The Higgs vacuum expectation value has given the electron a mass!  Using the known value of $v$ as determined from the $W$ boson mass, we can deduce the value of $y_e$ and hence the $h e^+ e^-$ Feynman rule (see Fig.~\ref{fig:hee}), which is
\begin{equation}
	h \bar e e: \quad \frac{-i y_e}{\sqrt{2}} = \frac{-i m_e}{v}.
	\label{eq:heefr}
\end{equation}
Thus the $h \bar e e$ coupling is uniquely predicted in the SM once the electron mass and $v$ are known.

The Higgs-electron coupling is really very small:
\begin{equation}
	\frac{y_e}{\sqrt{2}} = \frac{m_e}{v} 
	= \frac{511~{\rm keV}}{246~{\rm GeV}} \simeq 2.1 \times 10^{-6}.
\end{equation}
We can write down a similar Higgs coupling and mass term for the muon and for the tau lepton.  The tau Yukawa coupling is more ``respectable,'' though still kind of small:
\begin{equation}
	\frac{y_{\tau}}{\sqrt{2}} = \frac{m_{\tau}}{v} 
	= \frac{1.78~{\rm GeV}}{246~{\rm GeV}} \simeq 7.2 \times 10^{-3}.
\end{equation}
The SM does not provide any explanation for these numbers or their sizes; they are just parameters to be measured.  One can hope that a more complete theory of flavor would provide an explanation for the pattern of fermion masses.

Note that we have not generated any masses or Higgs couplings to neutrinos, because we did not introduce three right-handed neutrinos $\nu_R$ to participate in the Higgs couplings.  More on this after we deal with the quark masses.

\subsubsection{Quark masses and mixing}

We start by following our noses and writing a term just like for the charged leptons:
\begin{equation}
	\mathcal{L}_{\rm Yukawa} \supset - \left[ y_d \bar d_R \Phi^{\dagger} Q_L 
	+ y_d^* \bar Q_L \Phi d_R \right], 
\end{equation}
where again the second term is just the Hermitian conjugate of the first, and we will again assume that the dimensionless constant $y_d$ is real for now.  As for the leptons, we multiply out the SU(2)$_L$ doublets in unitarity gauge,
\begin{equation}
	\Phi^{\dagger} Q_L = \left( 0, \frac{v + h}{\sqrt{2}} \right) 
	\left( \begin{array}{c} u_L \\ d_L \end{array} \right) 
	= \frac{v + h}{\sqrt{2}} d_L,
\end{equation}
so that
\begin{equation}
	\mathcal{L}_{\rm Yukawa} \supset - \left( \frac{y_d v}{\sqrt{2}} \right) \bar d d 
	- \frac{y_d}{\sqrt{2}} h \bar d d.
\end{equation}
The first term is a mass for the down quark, $m_d = y_d v/\sqrt{2}$, and the second is an $h \bar d d$ coupling.

So far so good, but what about the up-type quark masses?  To generate these, we take advantage of a useful property of SU(2): the anti-doublet or ``conjugate'' doublet transforms in the same way as the doublet.\footnote{Contrast this to the case of SU(3), in which the anti-triplet does \emph{not} transform in the same way as the triplet.}  The conjugate Higgs doublet is given by
\begin{equation}
	\tilde \Phi \equiv i \sigma^2 \Phi^* 
	= i \left( \begin{array}{cc} 0 & -i \\ i & 0 \end{array} \right) 
	\left( \begin{array}{c} \phi^- \\ \phi^{0*} \end{array} \right)
	= \left( \begin{array}{c} \phi^{0*} \\ - \phi^- \end{array} \right),
\end{equation}
and has hypercharge $Y = -1/2$.  Using $\tilde \Phi$ we can write another gauge-invariant Lagrangian term,
\begin{equation}
	\mathcal{L}_{\rm Yukawa} \supset - \left[ y_u \bar u_R \tilde \Phi^{\dagger} Q_L
	+ y_u^* \bar Q_L \tilde \Phi u_R \right],
\end{equation}
where again the second term is the Hermitian conjugate of the first, and we will assume that the dimensionless constant $y_u$ is real for now.  Writing out the product of the SU(2)$_L$ doublets in unitarity gauge,
\begin{equation}
	\tilde \Phi^{\dagger} Q_L = \left( \frac{v + h}{\sqrt{2}}, 0 \right) 
	\left( \begin{array}{c} u_L \\ d_L \end{array} \right) 
	= \frac{v + h}{\sqrt{2}} u_L,
\end{equation}
so that
\begin{equation}
	\mathcal{L}_{\rm Yukawa} \supset - \left( \frac{y_u v}{\sqrt{2}} \right) \bar u u
	- \frac{y_u}{\sqrt{2}} h \bar u u.
\end{equation}
This is exactly what we need to describe the up-quark mass $m_u = y_u v/\sqrt{2}$ and its coupling to the Higgs.

This is fine if we want to describe a single generation of quarks.  But in the SM there are three generations of quarks!  We should really rewrite our left- and right-handed quark fields with a generation index $j$, 
\begin{equation}
	Q_{Lj}, \quad u_{Rj}, \quad d_{Rj}, \qquad j = 1,2,3.
\end{equation}
In general, we can write a gauge-invariant coupling of $Q_{L1}$ to a Higgs doublet and each of $u_{Rj}$ and $d_{Rj}$, with $j = 1,2,3$, and the same for $Q_{L2}$ and $Q_{L3}$.  The most general form of the quark Yukawa Lagrangian is
\begin{equation}
	\mathcal{L}_{\rm Yukawa}^q = - \sum_{i=1}^3 \sum_{j=1}^3
	\left[ y^u_{ij} \bar u_{Ri} \tilde \Phi^{\dagger} Q_{Lj}
	+ y^d_{ij} \bar d_{Ri} \Phi^{\dagger} Q_{Lj} \right] + {\rm h.c.},
	\label{eq:quarkyukawas}
\end{equation}
where h.c.\ stands for Hermitian conjugate.  The dimensionless couplings $y^u_{ij}$ and $y^d_{ij}$ are now the $(i,j)$ entries of 3$\times$3 complex matrices, containing a total of 18 complex coupling parameters!  Replacing $\Phi$ with its vacuum value $(0, v/\sqrt{2})^T$, we obtain the quark mass terms:
\begin{equation}
	\mathcal{L}_{\rm Yukawa}^q \supset 
	- \left( \bar u_1, \bar u_2, \bar u_3 \right)_R \mathcal{M}^u
	\left( \begin{array}{c} u_1 \\ u_2 \\ u_3 \end{array} \right)_L 
	- \left( \bar d_1, \bar d_2, \bar d_3 \right)_R \mathcal{M}^d
	\left( \begin{array}{c} d_1 \\ d_2 \\ d_3 \end{array} \right)_L + {\rm h.c.},
\end{equation}
where 
\begin{equation}
	\mathcal{M}^u_{ij} = \frac{v}{\sqrt{2}} y^u_{ij}, \qquad
	\mathcal{M}^d_{ij} = \frac{v}{\sqrt{2}} y^d_{ij}
\end{equation}
are the \emph{quark mass matrices} in generation space, each containing 9 complex entries.

We want to find the quark mass eigenstates.  To do that, we just need to diagonalize the two complex matrices $\mathcal{M}^u$ and $\mathcal{M}^d$.  Any such matrix can be transformed into a real diagonal matrix by multiplying it on the left and right by appropriate unitary transformation matrices.  We define four unitary matrices $U_L$, $U_R$, $D_L$, and $D_L$ according to
\begin{equation}
	\left( \begin{array}{c} u_1 \\ u_2 \\ u_3 \end{array} \right)_{L,R} 
	= U_{L,R} \left( \begin{array}{c} u \\ c \\ t \end{array} \right)_{L,R}, 
	\qquad
	\left( \begin{array}{c} d_1 \\ d_2 \\ d_3 \end{array} \right)_{L,R}
	= D_{L,R} \left( \begin{array}{c} d \\ s \\ b \end{array} \right)_{L,R},
\end{equation}
where $u,c,t,d,s,b$ are the quark mass eigenstates, such that\footnote{For a unitary matrix, $U^{-1} = U^{\dagger}$.}
\begin{equation}
	U_R^{-1} \mathcal{M}^u U_L = \left( \begin{array}{ccc} 
	m_u & 0 & 0 \\
	0 & m_c & 0 \\
	0 & 0 & m_t \end{array} \right), \qquad 
	D_R^{-1} \mathcal{M}^d D_L = \left( \begin{array}{ccc}
	m_d & 0 & 0 \\
	0 & m_s & 0 \\
	0 & 0 & m_b \end{array} \right).
\end{equation}
Note that diagonalizing the mass matrices $\mathcal{M}^u$ and $\mathcal{M}^d$ simultaneously diagonalizes the Yukawa matrices $y^u_{ij} = \frac{\sqrt{2}}{v} \mathcal{M}^u_{ij}$ and $y^d_{ij} = \frac{\sqrt{2}}{v} \mathcal{M}^d_{ij}$: this means that the Higgs couplings to $\bar q q$ are real and diagonal in the quark mass basis.  In particular, the Feynman rules are just
\begin{equation}
	h \bar q q: \quad \frac{-i y_q}{\sqrt{2}} = \frac{-i m_q}{v},
\end{equation}
where $y_q$ is the appropriate eigenvalue of the Yukawa matrix $y^u_{ij}$ or $y^d_{ij}$.

Notice that we've ``broken up'' the left-handed quark doublets by rotating the up-type quarks by $U_L$ and the down-type quarks by the different matrix $D_L$.  This shows up in the charged-current weak interactions, which change $u_{Lj} \leftrightarrow d_{Lj}$ within the \emph{same} (linear combination of) doublets.  Because the mass eigenstates of the down-type quarks are no longer matched up to the mass eigenstates of the up-type quarks, there are generation-changing weak interactions, which are described by the Cabibbo-Kobayashi-Maskawa (CKM) matrix.

In the charged-current interaction part of the Lagrangian we have the quark bilinears
\begin{equation}
	\bar u_{L1} \gamma^{\mu} d_{L1}, \qquad
	\bar u_{L2} \gamma^{\mu} d_{L2}, \qquad
	\bar u_{L3} \gamma^{\mu} d_{L3}.
\end{equation}
Their sum can be written in matrix form as
\begin{equation}
	J_L^{+\mu} = \left( \bar u_1, \bar u_2, \bar u_3 \right)_L \gamma^{\mu} 
	\left( \begin{array}{c} d_1 \\ d_2 \\ d_3 \end{array} \right)_L
	= \left( \bar u, \bar c, \bar t \right)_L U_L^{\dagger} \gamma^{\mu}
	D_L \left( \begin{array}{c} d \\ s \\ b \end{array} \right)_L
	= \left( \bar u, \bar c, \bar t \right)_L \gamma^{\mu}
	V \left( \begin{array}{c} d \\ s \\ b \end{array} \right)_L.
	\label{eq:chargedcurrent}
\end{equation}
The combination $U_L^{\dagger} D_L \equiv V$ is the CKM matrix.  Its elements are denoted by quark symbol subscripts; e.g., $V_{ud}$ is the $(1,1)$ element of $V$.  This indexing convention also helps one remember the form of Eq.~(\ref{eq:chargedcurrent}).

The CKM matrix is unitary:
\begin{equation}
	V^{\dagger} V = \left( U_L^{\dagger} D_L \right)^{\dagger} \left( U_L^{\dagger} D_L \right)
	= D_L^{\dagger} U_L U_L^{\dagger} D_L = 1.
\end{equation}
Note also that $U_R$ and $D_R$ have no physical consequences in the SM: $u_{Ri}$ and $d_{Ri}$ are not tied together in any way, so their relative basis rotations do not matter.

In the neutral current interactions, the photon couplings $Q$ and the $Z$ boson couplings $(T^3 - s_W^2 Q)$ are the same for each of the three generations.  The fermion bilinears involved in the neutral current can then be written out in generation space as, e.g.,
\begin{equation}
	\left( \bar u_1, \bar u_2, \bar u_3 \right)_L \gamma^{\mu}
	\left( \begin{array}{c} u_1 \\ u_2 \\ u_3 \end{array} \right)_L
	= \left( \bar u, \bar c, \bar t \right)_L U_L^{\dagger} \gamma^{\mu}
	U_L \left( \begin{array}{c} u \\ c \\ t \end{array} \right)_L
	= \left( \bar u, \bar c, \bar t \right)_L \gamma^{\mu}
	\left( \begin{array}{c} u \\ c \\ t \end{array} \right)_L.
\end{equation}
So the neutral currents are automatically \emph{flavor diagonal}, so long as the photon and $Z$ boson couplings to all three generations are \emph{universal}.  This is a manifestation of the GIM mechanism (after Glashow, Iliopoulos and Maiani~\cite{GIM}).  It is also why ``flavor changing neutral currents'' (FCNCs) provide such tight constraints on physics beyond the SM: they are absent at tree level in the SM, and the SM FCNCs induced at one-loop by $W$ boson exchange are typically quite small effects.

As a last comment, it is often convenient to work in the weak basis in which the up-type quarks are mass eigenstates.  The weak isospin doublets can then be written as
\begin{equation}
	\left( \begin{array}{c} u \\ d^{\prime} \end{array} \right)_L, \qquad
	\left( \begin{array}{c} c \\ s^{\prime} \end{array} \right)_L, \qquad
	\left( \begin{array}{c} t \\ b^{\prime} \end{array} \right)_L,
\end{equation}
where in generation space,
\begin{equation}
	\left( \begin{array}{c} d^{\prime} \\ s^{\prime} \\ b^{\prime} \end{array} \right)_L
	= V \left( \begin{array}{c} d \\ s \\ b \end{array} \right)_L.
\end{equation}

\subsubsection{An aside on neutrino masses}

If the neutrinos are Dirac particles (we do not know whether this is true; the other alternative is that they are Majorana particles, which are their own antiparticles), then we can introduce three right-handed neutrino fields $\nu_{Ri}$ ($i = 1,2,3$) and write Dirac neutrino masses in the same way as the up-type quark masses:
\begin{equation}
	\mathcal{L}_{\rm Yukawa} \supset - y_{\nu} \bar \nu_R \tilde \Phi^{\dagger} L_L + {\rm h.c.},
\end{equation}
or, including the charged lepton mass terms and the full three-generation structure [compare Eq.~(\ref{eq:quarkyukawas})],
\begin{equation}
	\mathcal{L}_{\rm Yukawa}^{\ell} = - \sum_{i=1}^3 \sum_{j=1}^3 \left[
	y^{\nu}_{ij} \bar \nu_{R_i} \tilde \Phi^{\dagger} L_{Lj}
	+ y^{\ell}_{ij} \bar e_{Ri} \Phi^{\dagger} L_{Lj} \right] + {\rm h.c.}
\end{equation}
Exactly as for the quarks, we get Dirac masses for the charged lepton mass eigenstates $e$, $\mu$, $\tau$ and the neutrino mass eigenstates $\nu_1$, $\nu_2$, $\nu_3$.  The weak isospin doublets can be written in the basis in which the charged leptons are mass eigenstates as
\begin{equation}
	\left( \begin{array}{c} \nu_e \\ e \end{array} \right)_L, \qquad
	\left( \begin{array}{c} \nu_{\mu} \\ \mu \end{array} \right)_L, \qquad
	\left( \begin{array}{c} \nu_{\tau} \\ \tau \end{array} \right)_L,
\end{equation}
where the ``flavor eigenstates'' of the neutrinos, $\nu_e$, $\nu_{\mu}$, and $\nu_{\tau}$, are related to their mass eigenstates by the lepton analogue of the CKM matrix, called the Maki-Nakagawa-Sakata-Pontecorvo (MNSP, or PMNS depending on your political affiliation) matrix $U$:
\begin{equation}
	\left( \begin{array}{c} \nu_e \\ \nu_{\mu} \\ \nu_{\tau} \end{array} \right)_L
	= U \left( \begin{array}{c} \nu_1 \\ \nu_2 \\ \nu_3 \end{array} \right)_L.
	\label{eq:MNSP}
\end{equation}
The elements of the MNSP matrix are denoted by indices as, e.g., $U_{e1}$ for the $(1,1)$ element.  This helps one remember the form of Eq.~(\ref{eq:MNSP}).

Note that the Yukawa couplings needed to generate the neutrino masses are extremely---some would say unreasonably---small: for a neutrino mass $m_{\nu} \sim 0.1$~eV, the corresponding neutrino Yukawa coupling would be
\begin{equation}
	\frac{y_{\nu}}{\sqrt{2}} = \frac{m_{\nu}}{v} \simeq 4 \times 10^{-13}.
\end{equation}

The other possibility for neutrinos is a ``Majorana mass.''  In terms of the SM fields, this is a term of the form $m \nu_L \nu_L$ (no bar!).  Neutrinos are the only known fermion for which we can construct a Majorana mass because they are electrically neutral, so that the Majorana mass term does not violate electric charge conservation.  Such a mass term is not gauge invariant under SU(2)$_L \times$U(1)$_Y$, but we can generate it after electroweak symmetry breaking by writing a more complicated term involving the Higgs field:\footnote{The Majorana mass term is more properly written as
\begin{equation}
	\mathcal{L}_{\rm Majorana} = - \frac{y^{\rm Maj}_{ij}}{\Lambda} \bar L_{Li}^c \tilde \Phi^* \tilde \Phi^{\dagger} L_{Lj},
\end{equation}
where the \emph{conjugate spinor} $\bar L_L^c \equiv - L_L^T C$, where $C = -i \gamma^2 \gamma^0$ is known as the charge conjugation matrix.  $\bar L_L^c$ transforms in the same way as a right-handed spinor $\bar \psi_R$ under the Lorentz group.  (I also included a generation-dependent prefactor $y^{\rm Maj}_{ij}$ to allow for different Majorana masses for the three generations.)  Majorana particles also show up in supersymmetry---in the Minimal Supersymmetric Standard Model, the gluinos and neutralinos are Majorana fermions.  A good reference for practical calculational techniques involving Majorana fermions in the familiar four-component spinor notation is Appendix A of Ref.~\cite{Haber:1984rc}.}
\begin{equation}
	\mathcal{L}_{\rm Majorana} = - \frac{(\tilde \Phi^{\dagger} L_L )^2}{\Lambda}.
\end{equation}
Counting up the dimensionality of the fields in the numerator of $\mathcal{L}_{\rm Majorana}$ quickly reveals that the field operator has dimension 5.  This is thus a \emph{nonrenormalizable} interaction, with coefficient $1/\Lambda$ where $\Lambda$ indicates the cutoff scale beyond which a more complete theory must reveal itself.

Such a term yields a neutrino mass $m_{\nu} = v^2/2 \Lambda$.  To get a neutrino mass of $m_{\nu} \sim 0.1$~eV requires $\Lambda \sim 3 \times 10^{14}$~GeV.  The more complete theory that yields the Majorana mass term usually involves a very heavy Majorana right-handed neutrino $\nu_R$ with mass of order the scale $\Lambda$.  This is known as the ``Type-I Seesaw.''

\subsubsection{CKM matrix parameter counting}

You may have heard that the CKM matrix (and also the MNSP matrix) can be specified by three angles and a phase.  Here's where that counting comes from.
\begin{itemize}
\item We start with a $3 \times 3$ complex matrix $V$: in general it contains 9 complex numbers, i.e., 18 independent real parameters.
\item $V$ is unitary, yielding 9 constraints of the form $V_{ab}^{\dagger} V_{bc} = \delta_{ac}$.  This leaves 9 independent real parameters.
\item We are free to absorb a phase out of $V$ into each left-handed field, by redefining $q_L \to e^{i \alpha_{q_L}} q_L$, with $q = u,d$ of each of the three generations.  This removes an arbitrary phase from each row or column of $V$.  But a common phase redefinition of all the $q_L$ has no effect on $V$, so this rephasing actually removes only $6-1 = 5$ unphysical phases.  This leaves $9-5 = 4$ physical free parameters in $V$.
\end{itemize}
To see that these four free parameters comprise three angles and a phase, note that a $3\times 3$ real unitary matrix---i.e., an orthogonal matrix---has three independent parameters (the familiar three Euler angles).  Thus $4-3 = 1$ of our CKM parameters must be a complex phase.  This phase is what gives rise to CP violation in the Standard Model weak interactions.\footnote{Note also that if we'd had only two generations, the CKM matrix would be fixed in terms of a single mixing angle and no phase.  The introduction of the CP-violating phase was part of the original motivation for Kobayashi and Maskawa to introduce the third generation~\cite{KM}.}

\subsection{Higgs self-couplings}

Finally let's return to the Higgs potential,
\begin{equation}
	\mathcal{L}_V = - V(\Phi) = \mu^2 \Phi^{\dagger} \Phi - \lambda (\Phi^{\dagger} \Phi)^2,
\end{equation}
and work out the self-interactions of the Higgs.  In unitarity gauge, 
\begin{equation}
	\Phi^{\dagger} \Phi = \frac{1}{2} (h + v)^2,
\end{equation}
and minimizing the potential gave us the relation $\mu^2 = \lambda v^2$, which we will use to eliminate $\mu^2$.  

\begin{figure}
\begin{center}
\includegraphics{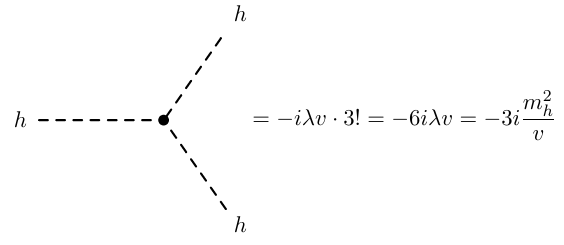}
\includegraphics{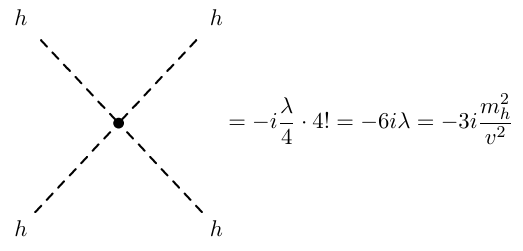}
\end{center}
\caption{Feynman rules for the $hhh$ and $hhhh$ vertices, as derived from the Lagrangian in Eq.~(\ref{eq:hhh}).  The $hhh$ coupling contains a symmetry factor of $3! = 6$ from the three identical Higgs bosons, and the $hhhh$ coupling contains a symmetry factor of $4! = 24$ from the four identical Higgs bosons.  See also Eqs.~(\ref{eq:hhhfr}) and (\ref{eq:hhhhfr}).}
\label{fig:hhh}
\end{figure}

Plugging in and multiplying out, we obtain
\begin{equation}
	\mathcal{L}_V = - \lambda v^2 h^2 - \lambda v h^3 - \frac{\lambda}{4} h^4 + {\rm const.}
	\label{eq:hhh}
\end{equation}
The first term is the mass term for the Higgs, $-\lambda v^2 = - m_h^2/2$.  The second term is an interaction vertex involving three Higgs bosons, with Feynman rule (see Fig.~\ref{fig:hhh})
\begin{equation}
	hhh: \quad - i \lambda v \times 3! = -6i \lambda v = - 3i \frac{m_h^2}{v},
	\label{eq:hhhfr}
\end{equation}
where the $3!$ is a combinatorical factor from the three identical Higgs bosons in the Lagrangian term.  The third term is an interaction vertex involving four Higgs bosons, with Feynman rule (see Fig.~\ref{fig:hhh})
\begin{equation}
	hhhh: \quad - i \frac{\lambda}{4} \times 4! = -6i \lambda = -3 i \frac{m_h^2}{v^2}, 
	\label{eq:hhhhfr}
\end{equation}
where again the $4!$ is a combinatorical factor from the four identical Higgs bosons in the Lagrangian term.

\section{SM Higgs collider phenomenology}
\label{sec:pheno}

All the masses of the SM particles\footnote{I'm ignoring neutrinos again, because they are irrelevant for Higgs phenomenology in the SM.} ($W^{\pm}$, $Z$, the charged fermions, and the Higgs as of summer 2012) are now known.  Therefore all the couplings of the Higgs boson relevant for Higgs collider phenomenology are uniquely predicted!  This means that any deviation from these predictions in Higgs phenomenology would provide evidence of physics beyond the SM.  (Before the Higgs discovery, $m_h$ was the only unknown parameter, and so the predictions were presented as a function of $m_h$.)

\subsection{Higgs decays}

Because we know the values of all the parameters that appear in the Higgs coupling Feynman rules, we can predict the partial widths for all the decays (and hence the decay branching ratios).  The SM predictions for these decay branching ratios are very important in the analysis of LHC Higgs data because they allow us to test the hypothesis that the discovered Higgs boson is the SM Higgs.  For that reason, a lot of work has been done to collect the most up-to-date calculations of the Higgs decay partial widths (including radiative corrections) and to make good estimates of their remaining theoretical uncertainties (from uncalculated higher-order radiative corrections) and parametric uncertainties (from uncertainties in the input parameters, like the quark masses).  At the time of writing, the most recent calculations and uncertainty estimates are summarized in Ref.~\cite{YR3}.

\subsubsection{$h \to f \bar f$}

\begin{figure}
\begin{center}
\includegraphics{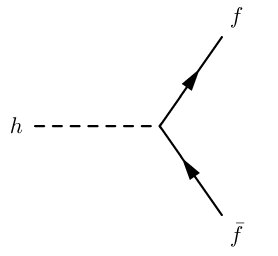}
\end{center}
\caption{Higgs boson decay to a fermion-antifermion pair.}
\label{fig:htoff}
\end{figure}

The Higgs boson can decay to a fermion-antifermion pair (see Fig.~\ref{fig:htoff}).  Because the Higgs-fermion interaction strength is proportional to the fermion mass, the decays to the heaviest kinematically-accessible fermion final states will have the largest partial widths.  Given the measured Higgs mass of about 125~GeV, decays to $t \bar t$ are way too off-shell to be numerically important.  Instead, the most important fermion final states are $b \bar b$, $\tau\tau$ and $c \bar c$.

The matrix element for the $h \to f \bar f$ process is
\begin{equation}
	i \mathcal{M} = \bar u_f \left( \frac{-i m_f}{v} \right) v_{\bar f},
\end{equation}
where $\bar u_f$ and $v_{\bar f}$ are the usual spinors for the outgoing fermion and antifermion.
Squaring the matrix element, summing over fermion polarizations (and colors, if applicable), and integrating over the final-state two-body phase space yields the partial width:
\begin{equation}
	\Gamma(h \to f \bar f) = \frac{N_c}{8 \pi} \frac{m_f^2}{v^2} m_h 
	\left[ 1 - \frac{4 m_f^2}{m_h^2} \right]^{3/2}.
	\label{eq:gammaff}
\end{equation}
This expression contains the following ingredients:
\begin{itemize}
\item It is proportional to the color factor $N_c = 3$ for quarks and $N_c = 1$ for leptons; this accounts for the sum over the three final-state colors in decays to quark-antiquark pairs.
\item It is proportional to the square of the Yukawa coupling $(m_f/v)$, as you would expect from looking at the matrix element.
\item It grows linearly with $m_h$: thus $\Gamma(h \to f \bar f)/m_h$, the width to $f \bar f$ as a fraction of the Higgs mass, would remain constant for fermionic decay modes if one were to crank up the Higgs mass.
\item It contains a kinematic factor $\left[ 1 - 4 m_f^2/m_h^2 \right]^{3/2} \equiv \beta^3$, which is $\simeq 1$ when the decay is well above threshold (i.e., when $m_h \gg 2 m_f$).\footnote{For decays to two equal-mass final-state particles, $\beta$ is just the speed in units of $c$ of either of the final-state particles in the parent particle's rest frame.  Well above threshold, the decay products are highly relativistic, so $\beta \simeq 1$.}
\end{itemize}
The state-of-the-art numerical predictions~\cite{YR3} for the partial widths for $h \to f \bar f$ contain more than just the leading-order expression in Eq.~(\ref{eq:gammaff}):
\begin{itemize}
\item The QCD corrections to $h \to q \bar q$ (i.e., decays to $b \bar b$ or $c \bar c$) are known to an astounding next-to-next-to-next-to-next-to-leading order (N$^4$LO).  The remaining uncertainty from uncalculated higher order QCD corrections is estimated to be only 0.1\%.\footnote{For $h \to q \bar q$, QCD corrections are quite significant and \emph{reduce} the partial width.  The dominant effect is captured by replacing $m_f$ in Eq.~(\ref{eq:gammaff}) by the running mass $\overline m_f(m_h)$ in the modified minimal subtraction ($\overline{\rm MS}$) renormalization scheme.}
\item The electroweak corrections are known at next-to-leading order (NLO).  The numerical results in Ref.~\cite{YR3} were computed using the public code {\tt HDECAY}~\cite{HDECAY}, which currently has the electroweak corrections implemented in a low-$m_h$ approximation leading to a 1--2\% uncertainty.
\item For $h \to b \bar b$, which constitutes the largest decay branching ratio for a 125~GeV SM Higgs boson and hence influences all the other decay branching ratios through its effect on the Higgs total width, there are parametric (input) uncertainties from $m_b$ and $\alpha_s$ (the latter contributes uncertainties in the QCD corrections).  The current uncertainty in the value of $m_b$ leads to a 3.3\% uncertainty in $\Gamma(h \to b \bar b)$ and the current uncertainty in the value of $\alpha_s$ leads to a 2.3\% uncertainty in $\Gamma(h \to b \bar b)$.
\end{itemize}
Few-percent-sized uncertainties are small compared to the uncertainties in the Higgs couplings after the first run of the LHC at 7--8~TeV.  However, with future plans to measure Higgs couplings at percent-level precision and below, we will have to find a way to reduce these theoretical and parametric uncertainties.  The parametric uncertainties in $\alpha_s$ and $m_b$ may be reduced a lot by advances in lattice QCD over the next five years~\cite{lattice}.

\subsubsection{$h \to W^+W^-$ and $h \to ZZ$}

\begin{figure}
\begin{center}
\includegraphics{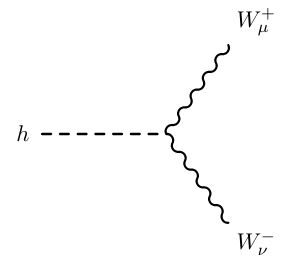}
\end{center}
\caption{Higgs boson decay to a $W^+W^-$ pair.}
\label{fig:htoww}
\end{figure}

Let's first consider how the Higgs decays to $W^+W^-$ and $ZZ$ (see Fig.~\ref{fig:htoww}) would behave in the case that the Higgs mass were high enough for these decays to be on-shell.  This turns out not to be the case in nature, but there's some nice physics to understand in the on-shell decay case.

The matrix element for $h \to W^+W^-$ is 
\begin{equation}
	i \mathcal{M} = 2 i \frac{M_W^2}{v} g^{\mu\nu} \epsilon^*_{\mu}(W^+) \epsilon_{\nu}(W^-),
\end{equation}
and similarly for $h \to ZZ$ replacing $M_W$ with $M_Z$.  Summing over the three polarization states of the massive $W$ bosons and integrating over the final-state phase space yields the decay width (for $m_h > 2 M_W$):
\begin{equation}
	\Gamma(h \to W^+W^-) = \frac{1}{16 \pi} \left( \frac{M_W^4}{v^2} \right) \frac{m_h^3}{M_W^4}
	\sqrt{1 - x_W} \left( 1 - x_W + \frac{3}{4} x_W^2 \right),
	\label{eq:GamWW}
\end{equation}
where $x_W \equiv 4 M_W^2/m_h^2$.  The expression for $\Gamma(h \to ZZ)$ is obtained by replacing $M_W$ with $M_Z$ everywhere and multiplying by an additional factor of $1/2$ to account for the fact that the two $Z$ bosons in the final state are identical.

This expression contains the following ingredients:
\begin{itemize}
\item It is proportional to the square of the coupling $M_W^2/v$, as you would expect from looking at the matrix element.
\item It grows with increasing Higgs mass like $m_h^3$: this can be traced back to the $E/M_W$ factors in the longitudinal $W$ boson polarization vectors.\footnote{In the $W$ rest frame the three polarization four-vectors $\epsilon^{\mu}$ are just $(0,1,0,0)$, $(0,0,1,0)$, and $(0,0,0,1)$.  Boosting along the $z$ axis such that the $W$ acquires a four-momentum $p^{\mu} = (E, 0, 0, p)$, the first two polarization vectors stay the same while the third (the longitudinal one) becomes $(p,0,0,E)/M_W$.}  
\item Another way to derive the $m_h$ dependence is via the Goldstone Boson Equivalence Theorem, which provides a good approximation for the partial width in the limit $m_h \gg 2 M_W$.  In this case the $W$ bosons can be replaced by the charged Goldstone bosons $\phi^{\pm}$.  The $h \phi^+ \phi^-$ vertex is proportional to $i \lambda v$, so $|\mathcal{M}| \sim \lambda^2 v^2$ and $\Gamma \sim \frac{1}{m_h} \lambda^2 v^2 \sim m_h^3/v^2$, where we have used the fact that $m_h^2 \sim \lambda v^2$.
\item The complicated dependence on the kinematic factor $x_W$ is due to the sum over the $W$ polarization vectors.  Note that the kinematic factor $\sqrt{1 - x_W} \left( 1 - x_W + \frac{3}{4} x_W^2 \right) \rightarrow 1$ when $m_h \gg 2 M_W$.
\end{itemize}

Of course, the expression in Eq.~(\ref{eq:GamWW}) is only valid for Higgs masses above the $WW$ threshold, which, since the Higgs discovery, we now know is not the case in nature.  Instead, one has to calculate off-shell $h \to WW^* \to Wf \bar f$ or the full doubly-offshell $h \to f \bar f f \bar f$.  This is a tedious calculation, but it has been done.  The current state-of-the-art theoretical predictions for SM Higgs decay to four fermions is implemented in a code called {\tt PROPHECY4F}~\cite{prophecy4f}, which includes NLO QCD and NLO electroweak corrections to the full $4f$ final states with all interferences included.  (For example, the processes $h \to W^*W^* \to \ell \nu \ell \nu$ and $h \to Z^*Z^* \to \ell \ell \nu \nu$ interfere with each other for same-flavor final state leptons.  This interference is important when $m_h < 2 M_W$ because the phase space overlap of the two processes becomes significant when the gauge bosons are forced off shell.)

The remaining theoretical uncertainty from missing higher-order radiative corrections is estimated to be only $\sim 0.5\%$.

\subsubsection{Loop-induced decays: $h \to gg$, $\gamma\gamma$, $Z\gamma$}

\begin{figure}
\begin{center}
\includegraphics{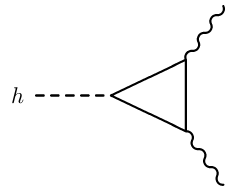}
\includegraphics{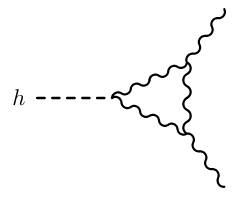}
\includegraphics{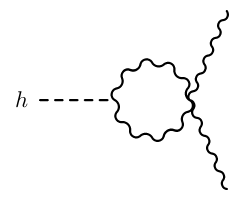}
\end{center}
\caption{Schematic Feynman diagrams for $h \to gg$ and $h \to \gamma\gamma$.}
\label{fig:loopdecays}
\end{figure}

The loop-induced Higgs decays are rare but important (remember that $h \to \gamma\gamma$ was one of the two Higgs boson discovery channels, along with $h \to ZZ^* \to 4\ell$).  Feynman diagrams are shown in Fig.~\ref{fig:loopdecays}.
\begin{itemize}
\item $h \to gg$: This decay is dominated by the top quark loop.  The bottom quark loop also contributes at the few-percent level.  
\item $h \to \gamma\gamma$: This decay is dominated by the $W$ boson loop.  The top quark loop contribution interferes destructively with the $W$ loop contribution, reducing the partial width by roughly 30\%.  The bottom quark and tau lepton loops also contribute a small amount.
\item $h \to Z \gamma$: This decay is dominated by the $W$ boson loop.  The top quark loop contribution is very small.
\end{itemize}

The partial width for Higgs decays to $\gamma\gamma$ takes the form~\cite{HHG}
\begin{equation}
	\Gamma(h \to \gamma\gamma) = \frac{\alpha^2}{256 \pi^3} \frac{m_h^3}{v^2}
	\left| \sum_i N_{ci} Q_i^2 F_i(\tau_i) \right|^2,
\end{equation}
where the sum runs over all particles that can run in the loop, $N_{ci}$ is the color factor for particle $i$ (3 for quarks, 1 for everything else), $Q_i$ is the electric charge of particle $i$ in units of $e$, and the loop factors $F_i$ for the $W$ boson (spin 1) and a fermion (spin 1/2) are given by
\begin{eqnarray}
	F_1(\tau) &=& 2 + 3 \tau + 3 \tau (2 - \tau) f(\tau), \nonumber \\
	F_{1/2}(\tau) &=& - 2 \tau \left[ 1 + (1 - \tau) f(\tau) \right],
\end{eqnarray}
where $\tau = 4 m_i^2/m_h^2$ and the function $f(\tau)$ is given by\footnote{Logarithms are always natural (i.e., log base $e$) in particle physics unless specified otherwise.}
\begin{equation}
	f(\tau) = \left\{ \begin{array}{ll}
	\left[ \sin^{-1} (1/\sqrt{\tau}) \right]^2 & {\rm for} \ m_h < 2 m_i, \\
	- \frac{1}{4} \left[ \log \left( \frac{1 + \sqrt{1 - \tau}}{1 - \sqrt{1 - \tau}} \right) - i \pi \right]^2
		& {\rm for} \ m_h > 2 m_i.
	\end{array} \right.
	\label{eq:f}
\end{equation}
The imaginary part in $f(\tau)$ for $m_h > 2 m_i$ is a consequence of the particles in the loop being kinematically able to go on shell.  When the particle in the loop is much heavier than the Higgs, $F_1 \to 7$ and $F_{1/2} \to -4/3$.  

\begin{figure}
\begin{center}
\includegraphics{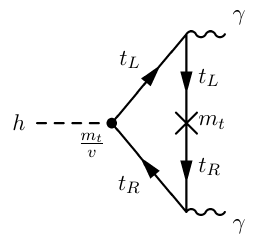}
\end{center}
\caption{One of the contributions to the fermion loop in $h \to \gamma\gamma$, showing the helicity of the fermions.  The helicity flip, which could be placed on any of the three fermion propagators, is marked with an $\times$.}
\label{fig:triangle}
\end{figure}

The $h \to \gamma\gamma$ partial width contains the following ingredients:
\begin{itemize}
\item The fermion loop asymptotes to a constant at large $m_f$.  This happens because the fermion triangle diagram generically goes like $1/m_f^2$ when the mass of the particle in the loop is large compared to any of the external invariant masses; however, this is multiplied by a factor of $m_f/v$ from the Yukawa coupling and another factor of $m_f$ from the fermion's helicity flip (see Fig.~\ref{fig:triangle}).
\item The contribution of light fermions (with masses $m_f \ll m_h$) to the amplitude falls like $m_f^2/m_h^2$ with decreasing fermion mass.
\item The $W$ boson loop amplitude asymptotes to a constant for $m_h \gg M_W$.  This limit can formally be obtained by taking $g \to 0$ while holding $v$ and $\lambda$ fixed.  In that case a contribution from the charged Goldstone boson running in the loop survives, leading to a finite amplitude dependent on $v$ and $\lambda$ (recall that $\lambda$ can be traded for $m_h$).
\item The partial width $\Gamma(h \to \gamma\gamma)$ grows with increasing Higgs mass like $m_h^3$.  This can be seen as a consequence of the effective-operator description of the interaction, which is required to take the form $h F^{\mu\nu} F_{\mu\nu}$ in order to preserve electromagnetic gauge invariance.  Expanding out the field strength tensors yields two factors of photon momentum in the amplitude.
\end{itemize}

The partial width for Higgs decays to $gg$ takes the form
\begin{equation}
	\Gamma(h \to gg) = \frac{\alpha_s^2}{256 \pi^3} \frac{m_h^3}{v^2} \times 2 
	\left| \sum_i F_{1/2}(\tau_i) \right|^2,
\end{equation}
where the factor of $2$ comes from the color structure and the sum runs over only the quarks, which are the only particles in the SM that both carry color and couple to the Higgs.

Finally, the partial width for Higgs decays to $Z \gamma$ can be written in a similar way in terms of a sum of amplitudes.  The loop functions take a different form because the final-state $Z$ boson has $p^2 = M_Z^2 \neq 0$, but the calculation is otherwise very similar to $h \to \gamma\gamma$.
The partial width takes the form
\begin{equation}
	\Gamma(h \to Z \gamma) = \frac{\alpha^2}{256 \pi^3} \frac{m_h^3}{v^2} \times 2
	\left| \sum_i A_i(\tau_i, \lambda_i) \right|^2 \left( 1 - \frac{M_Z^2}{m_h^2} \right)^3,
\end{equation}
where here the factor of $2$ comes from the fact that the two final-state gauge bosons are distinguishable (unlike in the case of $h \to \gamma\gamma$) and the last term in parentheses is a kinematic factor.
The loop factors $A_i$ for the $W$ boson and fermion loop contributions are given by~\cite{HHG}
\begin{eqnarray}
	A_W &=& -\cot\theta_W \left\{ \left[ 8 - \frac{16}{\lambda_W} \right] 
	I_2\left(\tau_W,\lambda_W\right)
	+ \left[\frac{4}{\lambda_W} \left(1+\frac{2}{\tau_W}\right) 
	- \left(6 + \frac{4}{\tau_W} \right) \right] I_1\left(\tau_W,\lambda_W\right)\right\}, \nonumber \\
	A_f &=& -4 N_{cf} Q_f 
	\frac{\left( \frac{1}{2}T^{3L}_f - Q_f s_W^2 \right)}{s_W c_W}
	\left[ I_1(\tau_f,\lambda_f)-I_2(\tau_f,\lambda_f) \right].
	\label{eq:Zgaamps}
\end{eqnarray}
Here $T^{3L}_f = \pm 1/2$ is the third component of isospin for the left-handed fermion $f$, so that $(\frac{1}{2} T^{3L}_f - Q_f s_W^2)/s_W c_W$ is the vectorial part of the $Z f \bar f$ coupling. 
The arguments of the functions are $\tau_i \equiv 4 m_i^2/m_h^2$ as before and $\lambda_i \equiv 4 m_i^2/M_Z^2$.

The loop factors are given in terms of the functions
\begin{eqnarray}
	 I_1(a,b) &=& \frac{ab}{2(a-b)} + \frac{a^2b^2}{2(a-b)^2} \left[f(a) - f(b)\right]
	 + \frac{a^2b}{(a-b)^2} \left[g(a) - g(b)\right], \nonumber \\
	 I_2(a,b) &=& -\frac{ab}{2(a-b)} \left[f(a) - f(b)\right],
\end{eqnarray}
where the function $f(\tau)$ was given in Eq.~(\ref{eq:f}) and
\begin{equation}
	g(\tau) = \left\{ \begin{array}{l l}
	\sqrt{\tau-1} \sin^{-1} \left(\sqrt{\frac{1}{\tau}}\right) & \quad  {\rm for} \ m_h < 2 m_i, \\
	\frac{1}{2} \sqrt{1-\tau} \left[ \log  \left( \frac{1 + \sqrt{1 - \tau}}{1 - \sqrt{1 - \tau}} \right) - i \pi \right] 
		& \quad  {\rm for} \ m_h > 2 m_i.
	\end{array} \right.
	\label{geq}
\end{equation}
Notice that, replacing the $ZWW$ and vectorial $Z f \bar f$ couplings with the corresponding $\gamma WW$ and $\gamma f \bar f$ couplings and taking $M_Z \to 0$ in the kinematic factors $\lambda_i$, one obtains $A_W \to - F_1(\tau_W)$ and $A_f \to - N_{cf} Q_f^2 F_{1/2}(\tau_f)$; that is, the amplitudes reduce exactly to the $h \to \gamma\gamma$ case, up to an overall minus sign that was built into the definitions of $A_W$ and $A_f$.

The current theoretical and parametric uncertainties in these loop-induced decays can be summarized as follows.
\begin{itemize}
\item $h \to gg$: The QCD corrections are known to N$^3$LO, leading to about a 3\% remaining scale uncertainty.\footnote{The ``scale uncertainty'' in QCD calculations is obtained by varying the renormalization scale by a factor of two in either direction about some chosen central value, and seeing how much the prediction changes as a consequence.  Because the renormalization scale dependence is an artifact of truncating the perturbation series at a finite order, the dependence of the result on the renormalization scale provides an estimate of how big the higher-order terms have to be in order to cancel this dependence once they are included.}  The electroweak corrections are known at NLO, leading to about a 1\% uncertainty from missing higher order corrections.  The current parametric uncertainty in $\alpha_s$ leads to an uncertainty in the $h \to gg$ partial width of about 4\%.\footnote{This comes from using $\alpha_s(M_Z) = 0.119 \pm 0.002$ (90\% CL) as advocated by the LHC Higgs Cross Section Working Group in 2012~\cite{YR2}.}
\item $h \to \gamma\gamma$: The QCD and electroweak corrections are each known to NLO, leaving a residual theoretical uncertainty of about 1\%.
\item $h \to Z \gamma$: This process has been calculated at leading order only.  The QCD corrections are expected to be small, since they affect only the fermion loop contributions which already give only a small contribution to the amplitude.  The uncertainty due to the missing NLO electroweak corrections is estimated at about 5\%.
\end{itemize}

\subsubsection{SM Higgs branching ratios}

To give some phenomenological insight, the SM Higgs branching ratios are summarized in order of size in Table~\ref{tab:BRs}.  All of the LHC measurements to date are roughly consistent with the SM predictions, within the current (large) uncertainties.

\begin{table}
\begin{center}
\begin{tabular}{c c l}
\hline\hline
Decay mode & BR & Notes (as of early 2014) \\
\hline
$b \bar b$ & 58\% & Observed at about $2\sigma$ at CMS \\
$WW^*$ & 22\% & Observed at $4\sigma$ \\
$gg$ & 8.6\% & \\
$\tau\tau$ & 6.3\% & Observed at 1--2~$\sigma$ \\
$c \bar c$ & 2.9\% & \\
$ZZ^*$ & 2.6\% & Discovery mode (in $ZZ^* \to 4\mu$, $2\mu 2e$, $4e$) \\
$\gamma\gamma$ & 0.23\% & Discovery mode \\
$Z\gamma$ & 0.15\% & \\
$\mu\mu$ & 0.022\% & \\
\hline \hline
$\Gamma_{\rm tot}$ & 4.1~MeV & \\
\hline\hline
\end{tabular}
\end{center}
\caption{Predicted decay branching ratios (BRs) for a 125~GeV SM Higgs boson, in order of size, from Ref.~\cite{YR2}.  The last row is the predicted Higgs total width.  Keep in mind that the relative uncertainties on the individual BR predictions are of order 3--10\%.}
\label{tab:BRs}
\end{table}

\subsection{Higgs production}

Because we know the values of all the parameters that appear in the Higgs coupling Feynman rules, we can also predict the cross sections for Higgs boson production in collisions of SM particles.  These cross sections are the second key ingredient (along with the Higgs branching ratios just discussed) in the analysis of Higgs data that allows us to test the hypothesis that the discovered Higgs boson is the SM Higgs.  As for the branching ratios, the current most up-to-date calculations and uncertainty estimates for Higgs production at the LHC are summarized in Ref.~\cite{YR3}.  I'll also comment on Higgs production in $e^+e^-$ collisions, relevant for Higgs studies at the proposed International Linear Collider (ILC).

\subsubsection{Higgs production in hadron collisions}

\begin{figure}
\begin{center}
\includegraphics{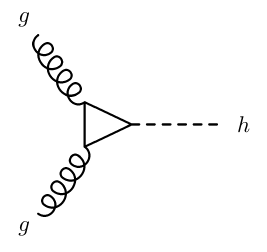} \hspace*{2cm}
\includegraphics{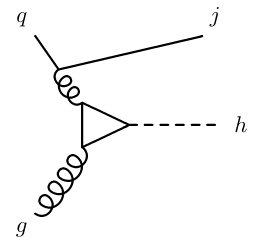}
\end{center}
\caption{Sample Feynman diagrams for Higgs boson production in gluon fusion, at leading order (left) and next-to-leading order (right).}
\label{fig:gf}
\end{figure}

The dominant Higgs production mode at the LHC is gluon fusion (abbreviated GF or ggF), $gg \to h$ (Fig.~\ref{fig:gf}).  This process makes up about 85\% of the total (inclusive) Higgs production cross section at the LHC.  At leading order, the amplitude for $gg \to h$ is the same as that for the decay $h \to gg$, with the initial and final states swapped.  (This amplitude must then be squared and integrated with the gluon parton densities.)  Beyond leading order, however, the QCD corrections for gluon-fusion Higgs production are different from those for the $h \to gg$ decay, because the additional radiated jets are in the final state, dramatically changing the kinematic structure.  These QCD corrections are quite large, enhancing the gluon-fusion Higgs production cross section by about a factor of two.   The current (2013), quite conservative, uncertainties on the gluon fusion Higgs production cross section at the 7--8~TeV LHC are about $\pm 8\%$ from QCD scale uncertainty and $\pm 7\%$ from the uncertainty in the parton distribution functions.  The Higgs discovery comes predominantly from this production mode.

\begin{figure}
\begin{center}
\includegraphics{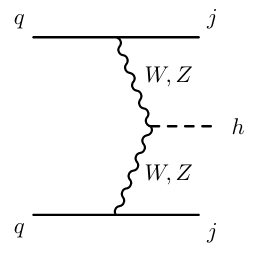}
\end{center}
\caption{Feynman diagram for weak boson fusion Higgs production at the LHC.}
\label{fig:VBF}
\end{figure}

The second-largest Higgs production cross section at the LHC is weak boson fusion (WBF), $qq \to hjj$, also known as vector boson fusion (VBF) (Fig.~\ref{fig:VBF}).  The cross section is about one tenth the size of that for gluon fusion.  The process is distinctive experimentally because the two incoming quarks tend to be scattered by only a small angle, leading to two very energetic jets pointing close to the beam line in opposite halves of the detector (referred to as ``forward jets'' or ``forward tagging jets'').  The process is interesting theoretically because it gives experimental access to the Higgs boson couplings to $WW$ and $ZZ$ in a production process.  VBF Higgs production has been seen at about the $2\sigma$ level in the $h\to \gamma\gamma$ final state.

\begin{figure}
\begin{center}
\includegraphics{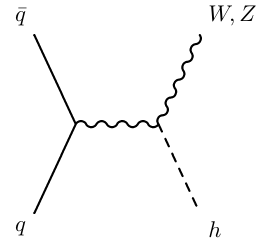}
\end{center}
\caption{Feynman diagram for associated production of a Higgs boson and a $W$ or $Z$ boson.}
\label{fig:Vh}
\end{figure}

Another distinctive Higgs production process is associated production together with a $W$ or $Z$ boson (Fig.~\ref{fig:Vh}).  The cross section for these two processes combined is about 60--70\% as large as that for VBF.  As for VBF, this process gives access to the Higgs boson coupling to $WW$ or $ZZ$.  Experimentally, the $W$ or $Z$ boson in the final state provides a useful handle to reduce background in searches for Higgs decays to $b \bar b$.

\begin{figure}
\begin{center}
\includegraphics{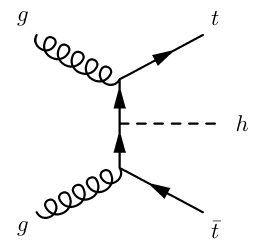}
\end{center}
\caption{Sample Feynman diagram for $t \bar t h$ associated production at the LHC.  There are also contributions in which the Higgs boson is attached to the outgoing top quark or antiquark line, as well as contributions from $q \bar q$ or $gg$ annihilation through an $s$-channel gluon.}
\label{fig:tth}
\end{figure}

A challenging but important process is $t \bar t h$ associated production, in which the Higgs boson is radiated off a top-antitop quark pair (Fig.~\ref{fig:tth}).  The production rate is very low at the 7--8~TeV LHC and still low at the 14~TeV LHC (a mere 1\% of the inclusive Higgs cross section at this higher energy), but this process is an essential probe of the Higgs boson coupling to top quarks.  Knowledge of the $t\bar t h$ coupling from a direct (tree-level) process like $t \bar t h$ is essential in order to probe for contributions to the loop-induced $ggh$ coupling from non-SM particles in the loop.

\begin{figure}
\begin{center}
\includegraphics{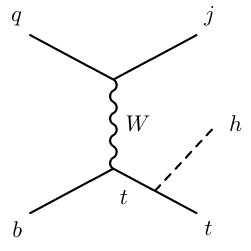} \hspace*{2cm}
\includegraphics{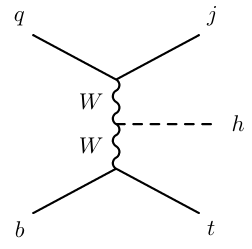}
\end{center}
\caption{Feynman diagrams for single-top plus Higgs production.}
\label{fig:th}
\end{figure}

Let me mention here a complementary (but even more experimentally challenging) process: single-top plus Higgs associated production.  This process gets contributions from two Feynman diagrams (Fig.~\ref{fig:th}), one in which the Higgs couples to the top quark and one in which it couples to the $W$ boson exchanged in the $t$-channel.  In the SM, there happens to be a strong destructive interference between these two diagrams, resulting in a cross section that is probably too small to be measured at the LHC.  However, it was pointed out recently~\cite{Farina:2012xp} that this process provides an interesting test of the \emph{relative sign} of the $WWh$ and $t \bar t h$ couplings, because a sign flip in one of the couplings relative to the SM would turn the destructive interference into constructive interference and make this cross section large enough to measure.

\begin{figure}
\begin{center}
\includegraphics{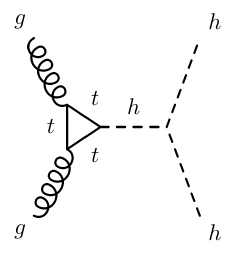} \hspace*{2cm}
\includegraphics{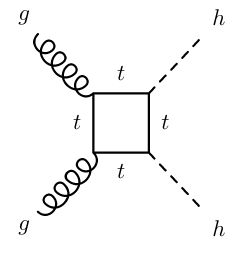}
\end{center}
\caption{Sample Feynman diagrams for double Higgs production in hadron collisions.}
\label{fig:doubleh}
\end{figure}

One last process worth studying at the LHC is double Higgs production (Fig.~\ref{fig:doubleh}).  The value of this process is that it allows an experimental probe of the Higgs self-coupling through the $hhh$ vertex.  The cross section is low and the process is experimentally very tough: current simulation studies indicate that one would need a high-luminosity run of the LHC (3000~fb$^{-1}$ at each of two detectors at 14~TeV) to get even a $\pm 30\%$ measurement of the triple-Higgs coupling $\lambda$~\cite{snowmass}.

\subsubsection{Higgs coupling extraction at the LHC}

Extracting the individual Higgs couplings from LHC data is a challenge because what's measured is the rates in individual production and decay channels.  In the zero-width approximation, a particular rate can be written as
\begin{equation}
	{\rm Rate}_{ij} = \sigma_i \times {\rm BR}_j = \sigma_i \times \frac{\Gamma_j}{\Gamma_{\rm tot}},
\end{equation}
where $\sigma_i$ depends on the production coupling, $\Gamma_j$ depends on the decay coupling, and $\Gamma_{\rm tot}$ depends on \emph{all} the couplings of the Higgs through which it can decay,
\begin{equation}
	\Gamma_{\rm tot} = \sum_k \Gamma_k.
\end{equation}
The total width depends on the couplings that control all of the most important Higgs decay modes.  It can also get a contribution from possible new non-SM decays that we don't know about.

The possibility of new, non-SM Higgs decays gives rise to a ``flat direction'' in the fit for the Higgs couplings based on Higgs production $\times$ decay rates.  The flat direction comes from adding a new component $\Gamma_{\rm new}$ to the Higgs total width while simultaneously cranking up all the SM Higgs couplings (so that the cross sections are increased).  Because there is no simple way at the LHC to measure any of the Higgs production couplings independent of branching ratios, this flat direction prevents the Higgs couplings from being extracted in a totally model independent way from Higgs data.  (Note that if $\Gamma_{\rm new}$ comprises decays to invisible particles such as dark matter candidates, it can be constrained through direct searches for invisibly-decaying Higgs events produced in VBF or $Wh/Zh$ associated production.  These searches are being done.  Here we are worried about decays that evade detection, such as into light-flavor jets.)

The simplest work-around is to do a constrained fit making use of a model assumption.  The approaches used in the literature are:
\begin{itemize}
\item Assume $\Gamma_{\rm new} = 0$, i.e., only SM decays are allowed.  This is valid if there are no new non-SM particles into which the Higgs can decay.
\item Assume that the $hWW$ and $hZZ$ couplings must be less than or equal to their SM values.  This is valid in extended Higgs models containing scalars that transform only as doublets and/or singlets of SU(2)$_L$.  It works because limiting these couplings prevents the VBF and associated $Wh/Zh$ production cross sections from being cranked up to accommodate $\Gamma_{\rm new}$.
\end{itemize}
In both cases, measurements in the VBF and associated $Wh/Zh$ production modes together with Higgs decays to $WW^*$ are important to constrain the $hWW$ coupling in both production and decay, and a measurement of Higgs decays to $b \bar b$ is important to constrain what is usually the single largest contribution to the Higgs total width.

It is worth mentioning here that, very recently, two clever new methods have been proposed to get a handle on the total width of the Higgs at the LHC by taking advantage of the relationship between $\Gamma_{\rm tot}$ and the Higgs production couplings along the flat direction:
\begin{itemize}
\item Interferometry in $gg \ (\to h) \to \gamma\gamma$: This method takes advantage of the interference between $gg \to h \to \gamma\gamma$ and the continuum $gg \to \gamma\gamma$ background, which produces a slight shift in the position of the Higgs mass peak in the $\gamma\gamma$ invariant mass distribution~\cite{Dixon}.  The size of the shift depends on the strength of the $hgg$ and $h \gamma\gamma$ effective couplings.  The shift could be measured by comparing the Higgs mass measurements in the $\gamma\gamma$ and $ZZ^* \to 4\ell$ final states, or by comparing the mass measurement in the $\gamma\gamma$ final state at low and high Higgs transverse momentum (the size of the interference effect is transverse momentum--dependent).  The sign of the mass shift provides access to the sign of the product of $hgg$ and $h \gamma\gamma$ couplings.  The sensitivity of this method is ultimately limited by the achievable precision in the Higgs mass measurement in the $\gamma\gamma$ final state.
\item Off-shell $gg \to h \to ZZ$: This method takes advantage of the fact that the Higgs coupling to $ZZ$ (and $WW$) is not small, and that the physical Higgs width is so small only because the Higgs mass is well below the $ZZ$ (and $WW$) threshold.  As a result, the cross section for $gg \to h^* \to ZZ$ through an off-shell Higgs boson is not totally negligible when the $ZZ$ invariant mass is above $2 M_Z$.  The size of this off-shell cross section depends on the strength of the $hgg$ and $hZZ$ couplings, and it can be directly measured through the $ZZ$ cross section as a function of the $ZZ$ invariant mass~\cite{Melnikov}.  The simplest interpretation of this measurement in terms of the Higgs production and decay couplings requires an assumption that there is not any additional new physics contributing to the $ZZ$ final state.
\end{itemize}
Both of these methods are expected to be limited to sensitivities of a few times the SM Higgs width.  Even though they formally eliminate the flat direction, they are not expected to be constraining enough to remove the need for model assumptions if one wants a high-precision Higgs coupling extraction from LHC data.

\subsubsection{Higgs production in $e^+e^-$ collisions}

Higgs production at an $e^+e^-$ collider such as the International Linear Collider (ILC)~\cite{ILC} would not provide the massive statistics of the LHC, but it is much cleaner in two senses.  First, all the backgrounds are electroweak in size, so hadronic decay modes like $h \to c \bar c$ and $h \to gg$ can actually be measured.  Second, the four-momentum of the initial state is known (up to ``beamstrahlung'' effects which smear the beam energies).

\begin{figure}
\begin{center}
\includegraphics{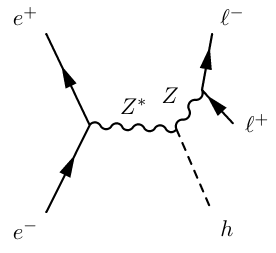}
\end{center}
\caption{$Zh$ associated production in $e^+e^-$ collisions.  Because the initial beam four-momenta are known, Higgs events can be selected by measuring the four-momenta of only the $Z$ decay products.}
\label{fig:recoilmass}
\end{figure}

The second feature allows for a key $e^+e^-$ technique known as the \emph{recoil mass method}, by which Higgs events can be selected by reconstructing the $Z$ boson without any reference to the Higgs decays.  To see how this works, define the four-momenta in Fig.~\ref{fig:recoilmass} as follows: the incoming electron and positron have momenta $p_1$ and $p_2$, respectively, the outgoing Higgs boson has momentum $k_1$, and the outgoing leptons $\ell^-$ and $\ell^+$ have momenta $k_2$ and $k_3$, respectively.  We can then define a Lorentz-invariant quantity that makes reference only to the four-momenta of the incident beams and the $Z$ decay products:
\begin{equation}
	M_{\rm rec}^2 \equiv (p_1 + p_2 - k_2 - k_3)^2.
\end{equation}
For the process $e^+e^- \to Zh$, conservation of four-momentum implies
\begin{equation}
	M_{\rm rec}^2 = k_1^2 = m_h^2.
\end{equation}
$Zh$ events thus appear as a bump in the $M_{\rm rec}$ distribution.  Crucially, this allows for a measurement of the cross section for $e^+e^- \to Zh$ without any reference to the Higgs branching ratios.  From this the $hZZ$ coupling can be directly measured, paving the way to a truly model-independent extraction of all the Higgs couplings.  One can also use the $Z$-tagged sample to measure the Higgs branching ratios directly and look for unexpected, non-SM decays.
For a Higgs mass of $125$~GeV, the cross section for $e^+e^- \to Zh$ peaks at an $e^+e^-$ center-of-mass energy of about 250~GeV.  This dictates the energy choice for the first phase of ILC running for Higgs studies.

\begin{figure}
\begin{center}
\includegraphics{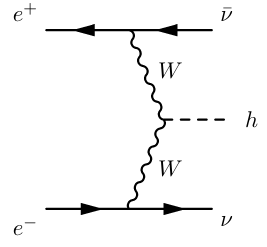} \hspace*{2cm}
\includegraphics{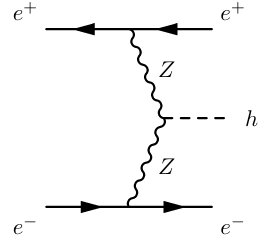}
\end{center}
\caption{Feynman diagrams for $WW$ fusion (left) and $ZZ$ fusion (right) in electron-positron collisions.}
\label{fig:eeVBF}
\end{figure}

\begin{figure}
\begin{center}
\includegraphics{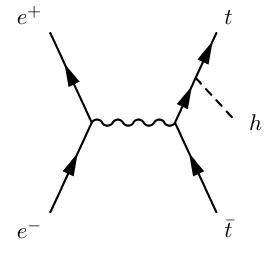}
\end{center}
\caption{One of the Feynman diagrams for $e^+e^- \to t \bar t h$.  (The other diagram has the Higgs attached to the $\bar t$ leg.)}
\label{fig:eetth}
\end{figure}

\begin{figure}
\begin{center}
\includegraphics{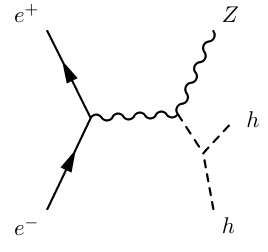} \hspace*{2cm}
\includegraphics{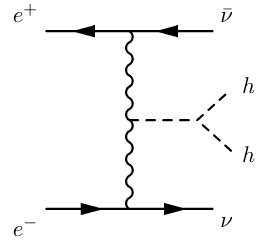}
\end{center}
\caption{Sample Feynman diagrams for double Higgs production in $e^+e^-$ collisions.}
\label{fig:eedoubleh}
\end{figure}

The other important Higgs production processes in $e^+e^-$ collisions are as follows:
\begin{itemize}
\item $WW$ fusion, Fig.~\ref{fig:eeVBF} (left).  The cross section for this process grows with center-of-mass energy.  This process provides most of the statistics for Higgs coupling measurements at higher center-of-mass energies (500~GeV and above).  The recoil mass technique cannot be used because of the two missing neutrinos.
\item $ZZ$ fusion, Fig.~\ref{fig:eeVBF} (right).  The cross section for this process is about a factor of 10 smaller than that for $WW$ fusion, mainly just due to the different strengths of the $W$ and $Z$ couplings to the electron lines.  This process does not contribute a lot to statistics, but on the other hand, the tagging electrons can be used to reconstruct the Higgs recoil mass.
\item $t \bar t h$ associated production, Fig.~\ref{fig:eetth}.  This process provides access to the $t \bar t h$ coupling.  The kinematic threshold is around 450~GeV, necessitating running at center-of-mass energies of at least $\sim 500$~GeV.
\item Double Higgs production, Fig.~\ref{fig:eedoubleh}.  This process provides access to the Higgs self-coupling.  As at the LHC, this is a very challenging measurement due to the small signal cross section.  With ILC running in a luminosity-upgraded machine configuration (i.e., the ``Lumi-Up'' option discussed in Ref.~\cite{snowmass}), it should be possible to measure $\lambda$ with around $\pm 13\%$ uncertainty.
\end{itemize}
Overall, the ILC program can be expected to provide model-independent measurements of the Higgs couplings to other SM particles with precisions in the few- to sub-percent range.  For a snapshot of the state of the field as of summer 2013, see the Snowmass Higgs Working Group report, Ref.~\cite{snowmass}.

\section{Higgs physics beyond the SM I: flavor (non-)conservation in a two-Higgs-doublet model}
\label{sec:2hdm}

The SM contains a number of features that happen ``by accident'' due to the simplicity of the SM, but often must be engineered ``by hand'' in extensions of the SM in order to be consistent with experimental constraints.  In this section and the next I illustrate two of these features of the SM---minimal flavor violation and custodial SU(2) symmetry---by exploring Higgs sector extensions that do not automatically preserve these features.

If the Higgs Yukawa couplings to fermions were absent, the SM would possess a global U(3)$^5$ flavor symmetry, comprising five different $3\times 3$ unitary transformations among the three generations of each type of chiral fermion (one for each of $Q_L$, $u_R$, $d_R$, $L_L$, and $e_R$).  In the SM, this large global flavor symmetry is explicitly broken by the Yukawa matrices $y^u_{ij}$, $y^d_{ij}$ and $y^{\ell}_{ij}$.  Extensions of the SM in which the global flavor symmetry is still broken \emph{only} by these three Yukawa matrices are said to have the property of \emph{minimal flavor violation}.  Models in this class tend to satisfy flavor constraints (e.g., kaon, $B$-meson, and $D$-meson oscillations and decays, $\mu \to e \gamma$, etc.) without too much tuning of parameters.  Models that contain new, non-minimal sources of flavor violation instead tend to be in gross violation of experimental constraints unless they are heavily tuned to evade the constraints.  This makes minimal flavor violation an attractive principle to implement in model-building.

To illustrate the consequences of non-minimal flavor violation, let's consider an extension of the SM containing two Higgs doublets---a two-Higgs-doublet model, or 2HDM.  Along the way I'll take the opportunity to illustrate some of the essential phenomenological features of 2HDMs.  We start with two copies of the SM Higgs doublet, with hypercharge $Y = 1/2$ as usual:
\begin{equation}
	\Phi_1 = \left( \begin{array}{c} \phi_1^+ \\ \phi_1^0 \end{array} \right), \qquad \qquad
	\Phi_2 = \left( \begin{array}{c} \phi_2^+ \\ \phi_2^0 \end{array} \right).
\end{equation}
In general, both of the Higgs doublets can carry a nonzero vev:
\begin{equation}
	\Phi_1 = \left( \begin{array}{c} \phi_1^+ \\ (h_1 + v_1 + i a_1)/\sqrt{2} \end{array} \right), \qquad \qquad
	\Phi_2 = \left( \begin{array}{c} \phi_2^+ \\ (h_2 + v_2 + i a_2)/\sqrt{2} \end{array} \right).
\end{equation}
Here we have already made two assumptions: 
\begin{itemize}
\item That $v_1$ and $v_2$ both lie in the neutral components of $\Phi_1$ and $\Phi_2$: this is essential to avoid breaking the gauge symmetry of electromagnetism!
\item That $v_1$ and $v_2$ are both real: this is an assumption that there is no CP violation in the scalar sector.  It is not strictly required by experimental constraints, but it makes things simpler.
\end{itemize}
Counting up the fields, we have two complex charged scalars $\phi_1^{\pm}$ and $\phi_2^{\pm}$, two CP-even real scalars $h_1$ and $h_2$, and two CP-odd real scalars $a_1$ and $a_2$.  Of these, one complex charged scalar and one CP-odd real scalar (in general linear combinations of the states listed above) are Goldstone bosons and can be gauged away.  Remaining in the spectrum are a single charged scalar $H^{\pm}$, two CP-even real scalars $h^0$ and $H^0$ (by convention, $h^0$ is the lighter one and $H^0$ is the heavier one), and one CP-odd real scalar $A^0$.

\subsection{Finding the Goldstone bosons}

\begin{figure}
\begin{center}
\includegraphics{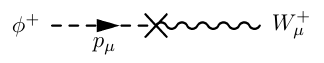}
\end{center}
\caption{The diagram corresponding to the $\partial_{\mu} \phi^+ W^{-\mu}$ term in Eq.~(\ref{eq:gaugegoldstone}).}
\label{fig:gaugegoldstone}
\end{figure}

First let's identify the linear combinations of the fields in $\Phi_1$ and $\Phi_2$ that are the Goldstone bosons (the physical charged and CP-odd scalars will then be the orthogonal combinations).  There are at least three ways to do this:
\begin{itemize}
\item Write out the full scalar potential, minimize it to find $v_1$ and $v_2$ in terms of the other parameters of the potential, isolate all terms quadratic in scalar fields, and diagonalize the resulting scalar mass-squared matrices.  The Goldstone bosons will be the massless eigenstates.
\item Apply the same SU(2)$_L \times$U(1)$_Y$ gauge transformations to $\Phi_1$ and $\Phi_2$ and determine which linear combinations of states can be entirely gauged away.
\item Identify the linear combinations of scalars that participate in Lagrangian terms of the form $\partial_{\mu} G^+ W^{- \mu}$.  This weird term (which disappears in unitarity gauge when the Goldstones are gauged away) describes a kind of mixing between the Goldstone boson and its corresponding gauge boson.
\end{itemize}
I find the third method particularly illuminating, so I'll follow it here.  Consider once more the gauge-kinetic terms involving the SM Higgs field:
\begin{equation}
	\mathcal{L} \supset (\mathcal{D}_{\mu} \Phi)^{\dagger} (\mathcal{D}^{\mu} \Phi),
\end{equation}
where 
\begin{equation}
	\mathcal{D}_{\mu} = \partial_{\mu} 
	- i \frac{g}{\sqrt{2}} (W^+_{\mu} \sigma^+ + W^-_{\mu} \sigma^-)
	- i \frac{e}{s_W c_W} Z_{\mu} (T^3 - s_W^2 Q) 
	- i e A_{\mu} Q
\end{equation}
and 
\begin{equation}
	\Phi = \left( \begin{array}{c} \phi^+ \\ (h + v + i a)/\sqrt{2} \end{array} \right),
\end{equation}
where the Higgs field is written in a general gauge (\emph{not} unitarity gauge).  Consider in particular the terms involving one $\partial_{\mu}$, one gauge field, one scalar, and one factor of $v$ (note that the $\partial_{\mu}$ must act on the scalar, because $\partial_{\mu} v = 0$ since $v$ is a constant):
\begin{eqnarray}
	\mathcal{L} &\supset& \left( \begin{array}{c} 
	\partial_{\mu} \phi^+ \\ (\partial_{\mu} h + i \partial_{\mu} a)/\sqrt{2} 
	\end{array} \right)^{\dagger}
	\left( \begin{array}{c}
	-i (g/\sqrt{2}) W^{+\mu} v \\ -i \frac{e}{s_W c_W} \left( -\frac{1}{2} \right) Z^{\mu} v
	\end{array} \right)
	+ \left( \begin{array}{c}
	-i (g/\sqrt{2}) W_{\mu}^+ v \\ -i \frac{e}{s_W c_W} \left( -\frac{1}{2} \right) Z_{\mu} v
	\end{array} \right)^{\dagger}
	\left( \begin{array}{c}
	\partial^{\mu} \phi^+ \\ (\partial^{\mu} h + i \partial^{\mu} a)/\sqrt{2}
	\end{array} \right) \nonumber \\
	&=& -i \frac{g}{\sqrt{2}} v \partial_{\mu} \phi^- W^{+\mu} 
	+ i \frac{g}{\sqrt{2}} v \partial_{\mu} \phi^+ W^{-\mu} 
	+ i \frac{e}{s_W c_W} \frac{v}{2} \frac{(\partial_{\mu} h - i \partial_{\mu} a)}{\sqrt{2}} Z^{\mu}
	- i \frac{e}{s_W c_W} \frac{v}{2} \frac{(\partial_{\mu} h + i \partial_{\mu} a)}{\sqrt{2}} Z^{\mu},
\end{eqnarray}
where in the first line the factors of $(-\frac{1}{2})$ come from the $T^3$ operator.  In the second line the $\partial_{\mu} h$ terms cancel and the $\partial_{\mu} a$ terms add, leading to (remember that the derivatives act only on the scalar field immediately to their right)
\begin{equation}
	\mathcal{L} \supset -i \frac{g}{\sqrt{2}} v \partial_{\mu} \phi^- W^{+\mu} 
	+ i \frac{g}{\sqrt{2}} v \partial_{\mu} \phi^+ W^{-\mu} 
	+ \frac{e}{\sqrt{2} s_W c_W} v \partial_{\mu} a Z^{\mu}.
	\label{eq:gaugegoldstone}
\end{equation}
These terms represent a weird two-particle ``vertex'' as shown in Fig.~\ref{fig:gaugegoldstone}.  This represents a kind of mixing of the Goldstone boson into its corresponding gauge boson.  These terms are eliminated in unitarity gauge and can be ignored as being unphysical.

Now let's do the same thing in the 2HDM:
\begin{equation}
	\mathcal{L} \supset -i \frac{g}{\sqrt{2}} \partial_{\mu} (v_1 \phi_1^- + v_2 \phi_2^-) W^{+ \mu}
	+ {\rm h.c.}
	+ \frac{e}{\sqrt{2} s_W c_W} \partial_{\mu} (v_1 a_1 + v_2 a_2) Z^{\mu}.
\end{equation}
The states involved in the unphysical interactions with the $W$ and $Z$ bosons (which \emph{must} be the Goldstones) are (after being properly normalized)
\begin{eqnarray}
	G^{\pm} &=& \frac{v_1}{\sqrt{v_1^2 + v_2^2}} \phi_1^{\pm}
	+ \frac{v_2}{\sqrt{v_1^2 + v_2^2}} \phi_2^{\pm}
	\equiv \cos \beta \, \phi_1^{\pm} + \sin \beta \, \phi_2^{\pm}, \nonumber \\
	G^0 &=& \frac{v_1}{\sqrt{v_1^2 + v_2^2}} a_1
	+ \frac{v_2}{\sqrt{v_1^2 + v_2^2}} a_2
	\equiv \cos \beta \, a_1 + \sin \beta \, a_2,
	\label{eq:goldstones}
\end{eqnarray}
where we have defined a mixing angle $\beta$ according to $\tan\beta \equiv v_2/v_1$.

This method of finding the Goldstone bosons always works and tends to be pretty easy to implement (certainly less work than minimizing the potential and finding the massless eigenstates).  It also makes clear that, even in crazy Higgs sector extensions containing your choice of representations of SU(2)$_L$, the identification of the Goldstone bosons depends only on the vevs of the scalars and possible group theoretic factors that arise from the covariant derivative acting on the scalar fields.

The orthogonal states, which are the physical charged and CP-odd mass eigenstates, now drop into our laps:
\begin{equation}
	H^{\pm} = - \sin\beta \, \phi_1^{\pm} + \cos\beta \, \phi_2^{\pm}, \qquad \qquad
	A^0 = - \sin\beta \, a_1 + \cos\beta \, a_2.
\end{equation}
$H^{\pm}$ must be a mass eigenstate because there are no other charged scalars in the theory for it to mix with.  $A^0$ can mix with $h_1$ and $h_2$ if CP is violated in the Higgs sector, but if CP is conserved it must be a mass eigenstate because there are no other CP-odd scalars for it to mix with.  In the latter case, $h_1$ and $h_2$ mix with each other to form two CP-even neutral scalar mass eigenstates:
\begin{equation}
	h^0 = - \sin\alpha \, h_1 + \cos\alpha \, h_2, \qquad \qquad
	H^0 = \cos\alpha \, h_1 + \sin\alpha \, h_2,
	\label{eq:hH}
\end{equation}
where the mixing angle $\alpha$ and the masses will be determined by the parameters of the scalar potential.

As we did in the SM, we now follow our noses and work out the $W$, $Z$, and fermion masses and their couplings to the Higgs bosons.

\subsection{Gauge boson mass generation}

The $W$ and $Z$ boson masses receive contributions from both Higgs doublets via the gauge-kinetic terms,
\begin{equation}
	\mathcal{L} \supset 
	\left( \mathcal{D}_{\mu} \Phi_1 \right)^{\dagger} \left( \mathcal{D}^{\mu} \Phi_1 \right)
	+ \left( \mathcal{D}_{\mu} \Phi_2 \right)^{\dagger} \left( \mathcal{D}^{\mu} \Phi_2 \right).
\end{equation}
The part of this expression involving only $h^0$, $H^0$, and the vevs is 
\begin{eqnarray}
	\mathcal{L} &\supset& 
	\frac{1}{2} (\partial_{\mu} h_1) (\partial^{\mu} h_1) 
	+ \frac{1}{2} (\partial_{\mu} h_2) (\partial^{\mu} h_2) \nonumber \\
	&& + \frac{1}{4} g^2 \left[ (h_1 + v_1)^2 + (h_2 + v_2)^2 \right] W^+_{\mu} W^{- \mu}
	\nonumber \\
	&& + \frac{1}{8} (g^2 + g^{\prime 2}) \left[ (h_1 + v_1)^2 + (h_2 + v_2)^2 \right] 
	Z_{\mu} Z^{\mu}.
	\label{eq:2hdmgauge}
\end{eqnarray}

From the first line of this expression, the unitary transformation from the $(h_1, h_2)$ basis to the $(h^0, H^0)$ basis gives the proper kinetic terms for the physical states, $\mathcal{L} \supset \frac{1}{2} (\partial_{\mu} h^0)(\partial^{\mu} h^0) + \frac{1}{2} (\partial_{\mu} H^0) (\partial^{\mu} H^0)$.

The masses of the $W$ and $Z$ come from the terms involving no scalar fields in the second and third lines of Eq.~(\ref{eq:2hdmgauge}):
\begin{eqnarray}
	M_W^2 &=& \frac{g^2}{4} (v_1^2 + v_2^2) = \frac{g^2 v_{\rm SM}^2}{4}, \nonumber \\
	M_Z^2 &=& \frac{g^2 + g^{\prime 2}}{4} (v_1^2 + v_2^2) = \frac{(g^2 + g^{\prime 2}) v_{\rm SM}^2}{4},
\end{eqnarray}
where the second equality in each line holds because we know that the $W$ and $Z$ masses are consistent with the SM prediction in terms of the SM Higgs vev $v_{\rm SM} \simeq 246$~GeV.  From this we find a constraint on the two vevs in the 2HDM, $v_1^2 + v_2^2 = v_{\rm SM}^2$.
	
The four-point couplings involving gauge bosons and CP-even neutral scalars also arise from the second and third lines of Eq.~(\ref{eq:2hdmgauge}):
\begin{eqnarray}
	\mathcal{L} &\supset & \frac{g^2}{4} \left[ h_1^2 + h_2^2 \right] W^+_{\mu} W^{- \mu}
	+ \frac{(g^2 + g^{\prime 2})}{8} \left[ h_1^2 + h_2^2 \right] Z_{\mu} Z^{\mu}
	\nonumber \\
	&=& \frac{g^2}{4} \left[ (h^0)^2 + (H^0)^2 \right] W^+_{\mu} W^{- \mu}
	+ \frac{(g^2 + g^{\prime 2})}{8} \left[ (h^0)^2 + (H^0)^2 \right] Z_{\mu} Z^{\mu}.
\end{eqnarray}
Each of these couplings is the same as the corresponding SM $hhVV$ coupling [compare Eqs.~(\ref{eq:Whcoups}) and~(\ref{eq:Zhcoups})].  This works out because the coefficients of the $h_1^2$ and $h_2^2$ terms are the same, so that the unitary rotation transforms $\left[ h_1^2 + h_2^2 \right]$ into $\left[ (h^0)^2 + (H^0)^2 \right]$ (similarly to the kinetic terms).

\begin{figure}
\begin{center}
\includegraphics{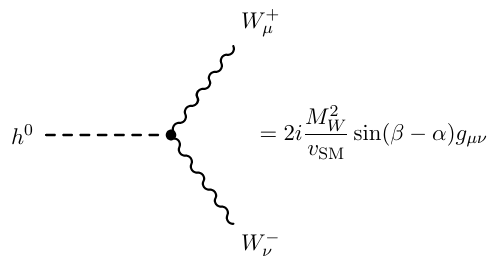} \hspace*{\fill}
\includegraphics{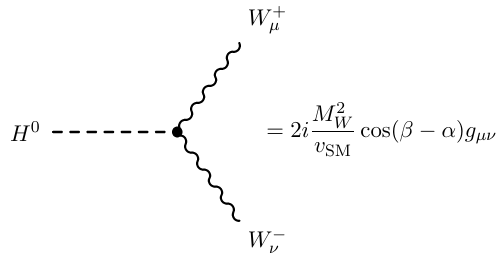} \\
\includegraphics{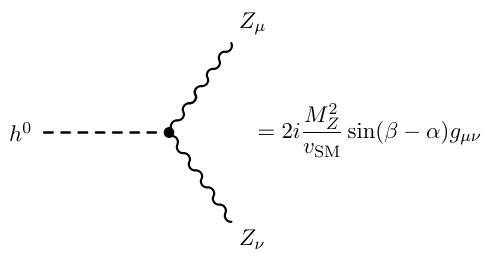} \hspace*{\fill}
\includegraphics{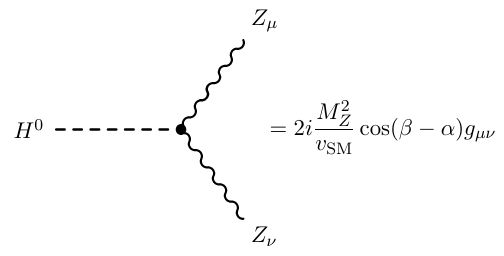}
\end{center}
\caption{Feynman rules for the couplings of $h^0$ and $H^0$ in two Higgs doublet models to $WW$ and $ZZ$ pairs.  See also Eq.~(\ref{eq:hvvfrs}).}
\label{fig:hvv2hdm}
\end{figure}

The three-point couplings involving one CP-even neutral scalar and two gauge bosons do not come out so simply.  From Eq.~(\ref{eq:2hdmgauge}) we have
\begin{equation}
	\mathcal{L} \supset \frac{g^2}{2} \left[ h_1 v_1 + h_2 v_2 \right] W^+_{\mu} W^{- \mu}
	+ \frac{(g^2 + g^{\prime 2})}{4} \left[ h_1 v_1 + h_2 v_2 \right] Z_{\mu} Z^{\mu}.
\end{equation}
Using
\begin{eqnarray}
	v_1 = v_{\rm SM} \cos\beta, &\qquad \qquad &
	v_2 = v_{\rm SM} \sin\beta, \nonumber \\
	h_1 = - \sin\alpha \, h^0 + \cos\alpha \, H^0, &\qquad \qquad &
	h_2 = \cos\alpha \, h^0 + \sin\alpha \, H^0,
\end{eqnarray}
we get
\begin{eqnarray}
	\left[ h_1 v_1 + h_2 v_2 \right] &=& 
	v_{\rm SM} h^0 (-\sin\alpha \cos\beta + \cos\alpha \sin\beta)
	+ v_{\rm SM} H^0 (\cos\alpha \cos\beta + \sin\alpha \sin\beta) \nonumber \\
	&=& v_{\rm SM} h^0 \sin(\beta - \alpha) + v_{\rm SM} H^0 \cos(\beta - \alpha).
\end{eqnarray}
The corresponding Feynman rules for $h^0$ and $H^0$ coupling to $WW$ and $ZZ$ are (see Fig.~\ref{fig:hvv2hdm})
\begin{eqnarray}
	h^0 W^+_{\mu} W^-_{\nu}: &\quad& 
	2 i \frac{M_W^2}{v_{\rm SM}} \sin(\beta - \alpha) g_{\mu\nu}, \nonumber \\
	h^0 Z_{\mu} Z_{\nu}: &\quad& 
	2 i \frac{M_Z^2}{v_{\rm SM}} \sin(\beta - \alpha) g_{\mu\nu}, \nonumber \\
	H^0 W^+_{\mu} W^-_{\nu}: &\quad& 
	2 i \frac{M_W^2}{v_{\rm SM}} \cos(\beta - \alpha) g_{\mu\nu}, \nonumber \\
	H^0 Z_{\mu} Z_{\nu}: &\quad& 
	2 i \frac{M_Z^2}{v_{\rm SM}} \cos(\beta - \alpha) g_{\mu\nu}.
	\label{eq:hvvfrs}
\end{eqnarray}
Each of these couplings is equal to the corresponding SM $hVV$ coupling except for the factor of $\sin(\beta - \alpha)$ or $\cos(\beta - \alpha)$.  We have uncovered a \emph{sum rule} that always applies in the 2HDM: the sum of the squares of the $h^0$ and $H^0$ couplings to $VV$ is equal to the square of the corresponding SM Higgs coupling to $VV$.

To see more clearly how this pattern of couplings arises, consider the situation when $(\beta - \alpha) = \pi/2$: then $\sin(\beta - \alpha) = 1$ (so that the $h^0VV$ coupling is equal to the corresponding coupling of the SM Higgs boson) and $\cos(\beta - \alpha) = 0$ (so that the $H^0VV$ coupling is equal to zero).  This situation corresponds to $\alpha = \beta - \pi/2$, so that
\begin{eqnarray}
	h^0 &=& - \sin\alpha \, h_1 + \cos\alpha \, h_2 \nonumber \\
	&=& - \sin\left(\beta - \frac{\pi}{2} \right) \, h_1 + \cos \left(\beta - \frac{\pi}{2} \right) \, h_2 \nonumber \\
	&=& \cos\beta \, h_1 + \sin\beta \, h_2.
\end{eqnarray}
Comparing this to the formulas for the Goldstone bosons in Eq.~(\ref{eq:goldstones}), and to 
\begin{equation}
	v_{\rm SM} = \sqrt{v_1^2 + v_2^2}
	= \frac{v_1^2}{\sqrt{v_1^2 + v_2^2}} + \frac{v_2^2}{\sqrt{v_1^2 + v_2^2}}
	= \cos\beta \, v_1 + \sin\beta \, v_2,
\end{equation}
we see that when $(\beta - \alpha) = \pi/2$, $h^0$ ``lives'' in the same linear combination of $\Phi_1$ and $\Phi_2$ as the Goldstones and the total vev!  In particular, we can rotate $\Phi_1$ and $\Phi_2$ by an angle $\beta$ to arrive at the ``Higgs basis,'' which for $(\beta - \alpha) = \pi/2$ reads
\begin{eqnarray}
	\Phi_H &\equiv & \cos\beta \, \Phi_1 + \sin\beta \, \Phi_2
	= \left( \begin{array}{c} G^+ \\ (h^0 + v_{\rm SM} + i G^0)/\sqrt{2} \end{array} \right), 
	\nonumber \\
	\Phi_0 &\equiv & -\sin\beta \, \Phi_1 + \cos\beta \, \Phi_2
	= \left( \begin{array}{c} H^+ \\ (H^0 + i A^0)/\sqrt{2} \end{array} \right).
\end{eqnarray}
The doublet $\Phi_0$ has zero vev, hence the subscript.  In this form, all the scalar fields are already written in terms of mass eigenstates, so the couplings to gauge bosons can be read off in a simple way by applying the covariant derivative.  Note, as usual for a unitary rotation, that
\begin{equation}
	\left( \mathcal{D}_{\mu} \Phi_1 \right)^{\dagger} \left( \mathcal{D}^{\mu} \Phi_1 \right)
	+ \left( \mathcal{D}_{\mu} \Phi_2 \right)^{\dagger} \left( \mathcal{D}^{\mu} \Phi_2 \right)
	= \left( \mathcal{D}_{\mu} \Phi_H \right)^{\dagger} \left( \mathcal{D}^{\mu} \Phi_H \right)
	+ \left( \mathcal{D}_{\mu} \Phi_0 \right)^{\dagger} \left( \mathcal{D}^{\mu} \Phi_0 \right).
\end{equation}

For $(\beta - \alpha) \neq \pi/2$ we move away from this idealized situation, but can still write the doublets in the Higgs basis (still defined so that the Goldstone bosons and all of the vev live in only one of the doublets):
\begin{eqnarray}
	\Phi_H &=& \left( \begin{array}{c} G^+ \\ 
	\left[h^0 \sin(\beta - \alpha) + H^0 \cos(\beta - \alpha) + v_{\rm SM} + i G^0\right]/\sqrt{2} 
	\end{array} \right), 
	\nonumber \\
	\Phi_0 &=& \left( \begin{array}{c} H^+ \\ 
	\left[- h^0 \cos(\beta - \alpha) + H^0 \sin(\beta - \alpha) + i A^0\right]/\sqrt{2} \end{array} \right).
\end{eqnarray}
This form makes the gauge couplings among physical states easy to read off.

\subsection{Fermion mass generation}

The most general gauge-invariant set of Yukawa couplings that we can write down involving two Higgs doublets is just two copies of the SM fermion mass generation terms:
\begin{eqnarray}
	\mathcal{L}_{\rm Yukawa} &=&
	- y^{\ell 1}_{ij} \bar e_{Ri} \Phi_1^{\dagger} L_{Lj}
	- y^{d1}_{ij} \bar d_{Ri} \Phi_1^{\dagger} Q_{Lj}
	- y^{u1}_{ij} \bar u_{Ri} \tilde \Phi_1^{\dagger} Q_{Lj} + {\rm h.c.}
	\nonumber \\
	&& - y^{\ell 2}_{ij} \bar e_{Ri} \Phi_2^{\dagger} L_{Lj}
	- y^{d2}_{ij} \bar d_{Ri} \Phi_2^{\dagger} Q_{Lj}
	- y^{u2}_{ij} \bar u_{Ri} \tilde \Phi_2^{\dagger} Q_{Lj} + {\rm h.c.},
	\label{eq:2hdmyukawas}
\end{eqnarray}
with an implicit sum over the generation indices $i,j = 1,2,3$.  Here the $y$ matrices are six general complex $3\times 3$ matrices of Yukawa couplings.

This form generically causes \emph{big trouble}!

To see why, let's look at the down-type quark mass terms:
\begin{eqnarray}
	\mathcal{L}_{\rm Yukawa} &\supset & 
	- \left( y^{d1}_{ij} \Phi_1^{\dagger} + y^{d2}_{ij} \Phi_2^{\dagger} \right) \bar d_{Ri} Q_{Lj}
	+ {\rm h.c.} \nonumber \\
	& \rightarrow & 
	- \left( y^{d1}_{ij} \frac{v_1}{\sqrt{2}} + y^{d2}_{ij} \frac{v_2}{\sqrt{2}} \right) \bar d_{Ri} d_{Lj}
	+ {\rm h.c.},
\end{eqnarray}
where in the second line we just keep the piece involving the vevs.  This means that the down-type quark mass matrix is
\begin{equation}
	\mathcal{M}^d_{ij} = \left( y^{d1}_{ij} \frac{v_1}{\sqrt{2}} + y^{d2}_{ij} \frac{v_2}{\sqrt{2}} \right).
\end{equation}
This is fine; it's just a general complex $3 \times 3$ matrix, which can be diagonalized in the same way as in the SM.  Diagonalizing $\mathcal{M}^d_{ij}$ diagonalizes the particular linear combination of $y^{d1}$ and $y^{d2}$ given by
\begin{equation}
	y^{d1}_{ij} \cos\beta + y^{d2}_{ij} \sin\beta,
\end{equation}
which is in fact the coefficient of the down-type quark coupling to $\Phi_H$ in the Higgs basis.

\begin{figure}
\begin{center}
\includegraphics{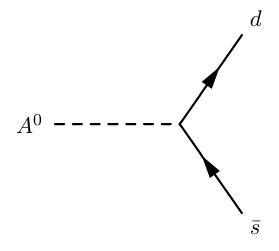}
\end{center}
\caption{A sample flavor-changing neutral Higgs coupling.}
\label{fig:Ads}
\end{figure}

\begin{figure}
\begin{center}
\includegraphics{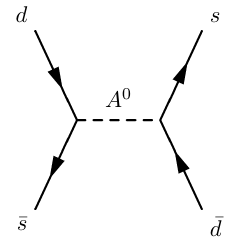}
\end{center}
\caption{A sample Feynman diagram for $K^0$--$\bar K^0$ mixing at tree level, mediated by a pseudoscalar Higgs boson $A^0$ with flavor-changing couplings to down-type quarks.}
\label{fig:KKbar}
\end{figure}

However, diagonalizing $\mathcal{M}^d_{ij}$ does \emph{not} in general diagonalize the orthogonal linear combination of $y^{d1}$ and $y^{d2}$,
\begin{equation}
	- y^{d1}_{ij} \sin\beta + y^{d2}_{ij} \cos\beta,
\end{equation}
which winds up being the coefficient of the down-type quark coupling to $\Phi_0$ in the Higgs basis.  When this linear combination is not diagonal in the mass basis, any neutral scalars that live in $\Phi_0$ will have \emph{flavor-changing} couplings to down-type fermions, such as the $A^0 \bar s d$ vertex shown in Fig.~\ref{fig:Ads}.  Such a coupling gives rise, e.g., to a tree-level contribution to $K^0$--$\bar K^0$ mixing, as shown in Fig.~\ref{fig:KKbar}.  These processes are known as \emph{flavor-changing neutral currents} and are generically an experimental disaster.  Similar problems happen in the up-type quark and lepton sectors.

The problem can be understood as being due to the breaking of the global flavor symmetry by more than one Yukawa matrix in each of the up-type quark, down-type quark, and charged lepton sectors.  This is \emph{not} minimal flavor violation.  

There are two well-known approaches to eliminate flavor-changing neutral Higgs couplings in 2HDMs by re-imposing minimal flavor violation.  The first is known as natural flavor conservation and was proposed as early as 1977 by Glashow and Weinberg and separately by Paschos~\cite{GWP}.  The second is called Yukawa alignment and represents the simplest implementation of a more general minimal flavor violating set-up.  The philosophies of the two approaches are different: in natural flavor conservation, the absence of flavor-changing neutral Higgs couplings is a consequence of a discrete symmetry obeyed by the Higgs sector itself, whereas in theories with Yukawa alignment, the flavor-changing neutral Higgs couplings are assumed to be prevented by the actions of an (unspecified) theory of flavor.

\subsection{Natural flavor conservation}

Natural flavor conservation is implemented by requiring that all fermions of a given electric charge are given their masses by only one Higgs doublet.  This is normally enforced using global $Z_2$ symmetries (parities), which may be softly broken in the scalar sector.  Under this condition, there can be only one $3 \times 3$ Yukawa matrix for up-type quarks, one for down-type quarks, and one for charged leptons.  Minimal flavor violation is thus achieved, and it is enforced by the Higgs sector itself.  Because the fermion mass matrices are then proportional to these Yukawa matrices, diagonalizing the fermion mass matrices diagonalizes all the Yukawa matrices, so that all neutral scalar couplings are flavor-diagonal in the fermion mass basis.  

In a 2HDM, natural flavor conservation is enforced by imposing a $Z_2$ symmetry under which one of the doublets transforms (e.g., $\Phi_2 \to - \Phi_2$ while $\Phi_1$ is unaffected) and under which some of the right-handed fermions transform (e.g., $u_{Ri} \to - u_{Ri}$ while all other fermion fields are unaffected).  In this example, the terms in Eq.~(\ref{eq:2hdmyukawas}) involving $\Phi_1$ and $\bar e_{Ri}$ or $\bar d_{Ri}$, and the term involving $\Phi_2$ and $\bar u_{Ri}$ are invariant under the $Z_2$ symmetry and hence allowed, while the other three terms are forbidden.

Ignoring neutrino masses, there are four physically distinct choices for the $Z_2$ charge assignments, leading to different phenomenology.  These choices define the four well-known ``types'' of 2HDMs.  In all cases we take $\Phi_2 \to - \Phi_2$ under the $Z_2$ symmetry while $\Phi_1 \to \Phi_1$.  The four types of 2HDM are then defined as follows:
\begin{itemize}
\item \emph{Type I:} $u_R, d_R, e_R \to -u_R, -d_R, -e_R$.  All right-handed fermions must couple to $\Phi_2$ and none to $\Phi_1$.  The Yukawa Lagrangian reads
\begin{equation}
	\mathcal{L}_{\Phi}^F =
	- y^{\ell 2}_{ij} \bar e_{Ri} \Phi_2^{\dagger} L_{Lj}
	- y^{d2}_{ij} \bar d_{Ri} \Phi_2^{\dagger} Q_{Lj}
	- y^{u2}_{ij} \bar u_{Ri} \tilde \Phi_2^{\dagger} Q_{Lj} + {\rm h.c.}
	\qquad {\rm (Type \ I)}.
\end{equation}
Inserting the vev of $\Phi_2$ in the usual way we acquire the fermion masses, 
\begin{equation}
	m_f = \frac{y_f v_2}{\sqrt{2}} = \frac{y_f v_{\rm SM}}{\sqrt{2}} \sin\beta,
\end{equation}
where $y_f$ is the appropriate eigenvalue of the appropriate Yukawa matrix.  The Yukawa coupling $y_f$ is then given by
\begin{equation}
	y_f = \frac{\sqrt{2} m_f}{v_{\rm SM}} \frac{1}{\sin\beta}.
\end{equation}
If we require that $y_t$ remain small enough that perturbative calculations involving this coupling remain reliable, we get a lower bound on $\sin\beta$; i.e., $v_2$ cannot be too small.  There is no corresponding lower bound on $v_1$, which could be taken all the way to zero without causing phenomenological trouble.  Taking into account the mixing angle $\alpha$ [recall Eq.~(\ref{eq:hH})], the Feynman rules for the $h^0 f \bar f$ and $H^0 f \bar f$ couplings are given by
\begin{eqnarray}
	h^0 f \bar f: &\quad& -i \frac{m_f}{v_{\rm SM}} \frac{\cos\alpha}{\sin\beta} 
	= -i \frac{m_f}{v_{\rm SM}} \left[ \sin(\beta - \alpha) + \cot\beta \cos(\beta - \alpha) \right],
	\nonumber \\
	H^0 f \bar f: &\quad& -i \frac{m_f}{v_{\rm SM}} \frac{\sin\alpha}{\sin\beta}
	= -i \frac{m_f}{v_{\rm SM}} \left[ -\cot\beta \sin(\beta - \alpha) + \cos(\beta - \alpha) \right].
\end{eqnarray}
Recall that the corresponding SM Higgs coupling to $f \bar f$ is $-i m_f/v_{\rm SM}$.  When $\sin(\beta - \alpha) \to 1$ (implying $\cos(\beta - \alpha) \to 0$), the $h^0 f \bar f$ coupling approaches that of the SM Higgs while the $H^0 f \bar f$ coupling becomes equal to the SM Higgs coupling times $-\cot\beta$.  Taking into account the perturbativity constraint on $\sin\beta$, which leads to $\cot\beta \lesssim \mathcal{O}(1)$, neither the $h^0$ nor $H^0$ coupling to fermions can be significantly enhanced over the corresponding SM Higgs coupling.

\item \emph{Type II:} $u_R \to -u_R$ and $d_R, e_R \to d_R, e_R$.  Right-handed up-type quarks must couple to $\Phi_2$ while right-handed down-type quarks and charged leptons must couple to $\Phi_1$ (note that this is the structure that appears in the Higgs sector of the Minimal Supersymmetric Standard Model).  The Yukawa Lagrangian reads
\begin{equation}
	\mathcal{L}_{\Phi}^F =
	- y^{\ell 1}_{ij} \bar e_{Ri} \Phi_1^{\dagger} L_{Lj}
	- y^{d1}_{ij} \bar d_{Ri} \Phi_1^{\dagger} Q_{Lj}
	- y^{u2}_{ij} \bar u_{Ri} \tilde \Phi_2^{\dagger} Q_{Lj} + {\rm h.c.}
	\qquad {\rm (Type \ II)}.
\end{equation}
Inserting the vevs we acquire the fermion masses,
\begin{equation}
	m_u = \frac{y_u v_2}{\sqrt{2}} = \frac{y_u v_{\rm SM}}{\sqrt{2}} \sin\beta, \qquad \qquad
	m_{d,\ell} = \frac{y_{d,\ell} v_1}{\sqrt{2}} = \frac{y_{d, \ell} v_{\rm SM}}{\sqrt{2}} \cos\beta,
\end{equation}
where again $y_{u,d,\ell}$ are the appropriate eigenvalues of the appropriate Yukawa matrices.  The Yukawa couplings $y_f$ are then given by
\begin{equation}
	y_u = \frac{\sqrt{2} m_u}{v_{\rm SM}} \frac{1}{\sin\beta}, \qquad \qquad
	y_{d, \ell} = \frac{\sqrt{2} m_{d, \ell}}{v_{\rm SM}} \frac{1}{\cos\beta}.
\end{equation}
If $v_2$ becomes too small, then $y_t$ becomes nonperturbatively large; similarly if $v_1$ becomes too small, then $y_b$ becomes nonperturbatively large.  This roughly constrains $0.5 \lesssim \tan\beta \lesssim 60$ (the exact choice of limits is a matter of taste).  The Feynman rules for the $h^0$ and $H^0$ couplings to fermions are then given by
\begin{eqnarray}
	h^0 u \bar u: &\quad& -i \frac{m_u}{v_{\rm SM}} \frac{\cos\alpha}{\sin\beta} 
	= -i \frac{m_u}{v_{\rm SM}} \left[ \sin(\beta - \alpha) + \cot\beta \cos(\beta - \alpha) \right],
	\nonumber \\
	h^0 d \bar d: &\quad& -i \frac{m_d}{v_{\rm SM}} \frac{-\sin\alpha}{\cos\beta}
	= -i \frac{m_d}{v_{\rm SM}} \left[ \sin(\beta - \alpha) - \tan\beta \cos(\beta - \alpha) \right],
	\nonumber \\
	H^0 u \bar u: &\quad& -i \frac{m_u}{v_{\rm SM}} \frac{\sin\alpha}{\sin\beta}
	= -i \frac{m_f}{v_{\rm SM}} \left[ -\cot\beta \sin(\beta - \alpha) + \cos(\beta - \alpha) \right],
	\nonumber \\
	H^0 d \bar d: &\quad& -i \frac{m_d}{v_{\rm SM}} \frac{\cos\alpha}{\cos\beta}
	= -i \frac{m_d}{v_{\rm SM}} \left[ \tan\beta \sin(\beta - \alpha) + \cos(\beta - \alpha) \right].
\end{eqnarray}
The couplings to leptons are obtained from the down-type quark couplings by replacing $m_d \to m_{\ell}$.  As before, in the limit $\sin(\beta - \alpha) \to 1$, the couplings of $h^0$ reduce to those of the SM Higgs boson.  Notice that, while the scalar couplings to up-type quarks cannot be significantly enhanced compared to the corresponding SM Higgs couplings, the scalar couplings to down-type quarks and charged leptons contain a factor of $\tan\beta$ and can be quite significantly enhanced.  This feature plays an important role in the phenomenology of the Type II 2HDM when $\tan\beta$ is large.

\item \emph{Lepton-specific} or \emph{Type X:} $u_R, d_R \to -u_R, -d_R$ and $e_R \to e_R$.  All right-handed quarks must couple to $\Phi_2$ while right-handed charged leptons must couple to $\Phi_1$.  The Yukawa Lagrangian reads
\begin{equation}
	\mathcal{L}_{\Phi}^F =
	- y^{\ell 1}_{ij} \bar e_{Ri} \Phi_1^{\dagger} L_{Lj}
	- y^{d2}_{ij} \bar d_{Ri} \Phi_2^{\dagger} Q_{Lj}
	- y^{u2}_{ij} \bar u_{Ri} \tilde \Phi_2^{\dagger} Q_{Lj} + {\rm h.c.}
	\qquad {\rm (Lepton \ specific)}.
\end{equation}
The Higgs couplings to quarks are the same as in the Type I 2HDM, while the couplings to leptons are given by
\begin{eqnarray}
	h^0 \ell \bar \ell: &\quad& -i \frac{m_{\ell}}{v_{\rm SM}} \frac{-\sin\alpha}{\cos\beta}
	= -i \frac{m_{\ell}}{v_{\rm SM}} \left[ \sin(\beta - \alpha) - \tan\beta \cos(\beta - \alpha) \right],
	\nonumber \\
	H^0 \ell \bar \ell: &\quad& -i \frac{m_{\ell}}{v_{\rm SM}} \frac{\cos\alpha}{\cos\beta}
	= -i \frac{m_{\ell}}{v_{\rm SM}} \left[ \tan\beta \sin(\beta - \alpha) + \cos(\beta - \alpha) \right].
\end{eqnarray}
In this model $\tan\beta$ can be large as 200 before the tau Yukawa coupling becomes nonperturbative, so that the scalar couplings to leptons can be quite significantly enhanced.

\item \emph{Flipped} or \emph{Type Y:} $u_R, e_R \to -u_R, -e_R$ and $d_R \to d_R$.  Right-handed up-type quarks and charged leptons must couple to $\Phi_2$ while right-handed down-type quarks must couple to $\Phi_1$.  The Yukawa Lagrangian reads
\begin{equation}
	\mathcal{L}_{\Phi}^F =
	- y^{\ell 2}_{ij} \bar e_{Ri} \Phi_2^{\dagger} L_{Lj}
	- y^{d1}_{ij} \bar d_{Ri} \Phi_1^{\dagger} Q_{Lj}
	- y^{u2}_{ij} \bar u_{Ri} \tilde \Phi_2^{\dagger} Q_{Lj} + {\rm h.c.}
	\qquad {\rm (Flipped)}.
\end{equation}
The Higgs couplings to quarks are the same as in the Type II 2HDM, while the couplings to leptons are given by 
\begin{eqnarray}
	h^0 \ell \bar \ell: &\quad& -i \frac{m_{\ell}}{v_{\rm SM}} \frac{\cos\alpha}{\sin\beta} 
	= -i \frac{m_{\ell}}{v_{\rm SM}} \left[ \sin(\beta - \alpha) + \cot\beta \cos(\beta - \alpha) \right],
	\nonumber \\
	H^0 \ell \bar \ell: &\quad& -i \frac{m_{\ell}}{v_{\rm SM}} \frac{\sin\alpha}{\sin\beta}
	= -i \frac{m_{\ell}}{v_{\rm SM}} \left[ -\cot\beta \sin(\beta - \alpha) + \cos(\beta - \alpha) \right].
\end{eqnarray}
The perturbativity constraints on $\tan\beta$ are the same as in the Type II 2HDM.  While the scalar couplings to down-type quarks can be quite significantly enhanced as in the Type II model, the scalar couplings to charged leptons contain a factor of $\cot\beta$ and cannot be significantly enhanced compared to the corresponding SM Higgs couplings.
\end{itemize}
Each of these models has its own distinctive phenomenology, which has been explored in great detail in the literature.

\subsection{Yukawa alignment}

Yukawa alignment is the simplest implementation of the full general minimal flavor violation framework in a model with more than one Higgs doublet.  It is implemented by allowing both doublets to couple to all fermions as in Eq.~(\ref{eq:2hdmyukawas}), but requiring that the two Yukawa matrices for each type of fermion are proportional to each other:
\begin{equation}
	y^{\ell 1}_{ij} = z_{\ell} y^{\ell 2}_{ij}, \qquad
	y^{d1}_{ij} = z_d y^{d2}_{ij}, \qquad
	y^{u1}_{ij} = z_u y^{u2}_{ij},
	\label{eq:alignment}
\end{equation}
where $z_{\ell}$, $z_d$, and $z_u$ are three free parameters, which can be complex in general.  By rotating to the Higgs basis, the parameter $\tan\beta$ can be absorbed into the definitions of $z_{\ell, d, u}$.

Minimal flavor violation is assumed to be enforced by some unspecified model of flavor outside of the Higgs sector itself, which gives rise to the proportionalities in Eq.~(\ref{eq:alignment}).  This structure yields more parameter freedom than the four ``types'' of natural-flavor-conserving 2HDMs discussed in the previous section, and in fact can be used to interpolate continuously between the ``types''.

\section{Higgs physics beyond the SM II: custodial symmetry and models with Higgs triplets}
\label{sec:triplets}

\subsection{The $\rho$ parameter and custodial symmetry}

The $\rho$ parameter was introduced to describe the relative strength of neutral-current and charged-current weak interaction processes at four-momentum transfers much smaller than the masses of the $W$ and $Z$ bosons.

\begin{figure}
\begin{center}
\includegraphics{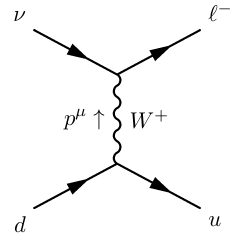}
\end{center}
\caption{A charged-current scattering process, $\nu d \to \ell^- u$.}
\label{fig:cc}
\end{figure}

Consider low-energy processes mediated by $W$ exchange at small momentum transfers, for example the charged-current neutrino scattering process $\nu d \to \ell^- u$ (Fig.~\ref{fig:cc}).  When $p^2 \ll M_W^2$, the $W$ propagator can be approximated by neglecting $p^2$:
\begin{equation}
	\frac{-i g^{\mu\nu}}{p^2 - M_W^2} \quad \rightarrow \quad \frac{i g^{\mu\nu}}{M_W^2}.
\end{equation}
This allows us to describe the process by a low-energy effective Lagrangian,
\begin{eqnarray}
	i \Delta \mathcal{L}_W &=& \frac{ig}{\sqrt{2}} (\bar e_L \gamma_{\mu} \nu_L) 
	\times \frac{i g^{\mu\nu}}{M_W^2} 
	\times \frac{ig}{\sqrt{2}} (\bar u_L \gamma_{\nu} d_L) \nonumber \\
	&=& \frac{-ig^2}{M_W^2} 
	\times \frac{1}{\sqrt{2}} (\bar e_L \gamma_{\mu} \nu_L) 
	\times \frac{1}{\sqrt{2}} (\bar u_L \gamma^{\mu} d_L) \nonumber \\
	&\equiv & \frac{-i g^2}{M_W^2} J^{-}_{W\mu} J^{+ \mu}_W,
\end{eqnarray}
where $J^{\pm}_{W\mu}$ are ``charged currents'' in analogy to the electromagnetic current.  This is Fermi's original four-fermi(on) theory, with
\begin{equation}
	\frac{g^2}{M_W^2} = \frac{8 G_F}{\sqrt{2}} = \frac{4}{v^2},
\end{equation}
so that 
\begin{equation}
	\Delta \mathcal{L}_W = -\frac{8 G_F}{\sqrt{2}} 
	\times \frac{1}{\sqrt{2}} (\bar e_L \gamma_{\mu} \nu_L) 
	\times \frac{1}{\sqrt{2}} (\bar u_L \gamma^{\mu} d_L).
\end{equation}

\begin{figure}
\begin{center}
\includegraphics{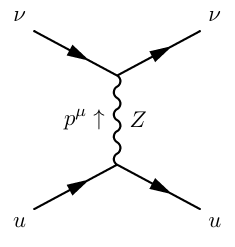}
\end{center}
\caption{A neutral-current scattering process, $\nu u \to \nu u$.}
\label{fig:nc}
\end{figure}

The same formalism can be used to describe the \emph{weak neutral currents}\footnote{The weak neutral currents were predicted in 1973 in the Glashow-Weinberg-Salam SU(2)$_L \times$U(1)$_Y$ Standard Model.  Their existence was experimentally confirmed in 1974 at CERN in a neutrino scattering experiment.} mediated by $Z$ boson exchange at small momentum transfers, for example the neutral-current neutrino scattering process $\nu u \to \nu u$ (Fig.~\ref{fig:nc}).  When $p^2 \ll M_Z^2$, the $Z$ propagator can be approximated by neglecting $p^2$:
\begin{equation}
	\frac{-i g^{\mu\nu}}{p^2 - M_Z^2} \quad \rightarrow \quad \frac{i g^{\mu\nu}}{M_Z^2}.
\end{equation}
This allows us to describe the process by a low-energy effective Lagrangian,
\begin{eqnarray}
	i \Delta \mathcal{L}_Z &=& \frac{-i g}{c_W} \left( \bar \nu \gamma_{\mu} (T^3 - s_W^2 Q) \nu \right)
	\times \frac{i g^{\mu\nu}}{M_Z^2} 
	\times \frac{-i g}{c_W} \left( \bar u \gamma_{\nu} (T^3 - s_W^2 Q) u \right)
	\nonumber \\
	&=& \frac{-i g^2}{c_W^2 M_Z^2} 
	\times \left( \bar \nu \gamma_{\mu} (T^3 - s_W^2 Q) \nu \right)
	\times \left( \bar u \gamma_{\nu} (T^3 - s_W^2 Q) u \right) \nonumber \\
	&\equiv & -i \frac{8 G_F}{\sqrt{2}} \rho  
	\times \left( \bar \nu \gamma_{\mu} (T^3 - s_W^2 Q) \nu \right)
	\times \left( \bar u \gamma_{\nu} (T^3 - s_W^2 Q) u \right).
\end{eqnarray}
Here the ``neutral currents'' $J^0_{Z\mu}$ can be defined as the fermion bilinears in the parentheses in analogy to the charged currents $J^{\pm}_{W\mu}$.  Note the appearance of the parameter $\rho$, called the ``rho parameter,'' which is \emph{defined} as the ratio of strengths of the neutral to charged currents:
\begin{equation}
	\rho = \left( \frac{g^2}{c_W^2 M_Z^2} \right) \times \left(\frac{g^2}{M_W^2} \right)^{-1}
	= \frac{M_W^2}{c_W^2 M_Z^2}.
	\label{eq:rho}
\end{equation}
In particular, the factor of $c_W^2$ in this expression comes from the coupling strength of the $Z$ boson, which originated from the SU(2)$_L$ and U(1)$_Y$ gauge couplings combined with the $\gamma$--$Z$ mixing angle.

But in the SM, we found that the $W$ and $Z$ masses are predicted to be
\begin{equation}
	M_W^2 = \frac{g^2 v^2}{4}, \qquad \qquad
	M_Z^2 = \frac{(g^2 + g^{\prime 2}) v^2}{4} = \frac{g^2 v^2}{4 c_W^2},
\end{equation}
where now the $c_W^2$ in the expression for $M_Z^2$ comes from the gauge boson mass-squared matrix.  Plugging these expressions for $M_W^2$ and $M_Z^2$ into Eq.~(\ref{eq:rho}) for the rho parameter, we find 
\begin{equation}
	\rho = 1
\end{equation} 
at tree level in the SM.

At this point the reader's natural reaction is ``yeah, so what?''  In fact, $\rho = 1$ is a consequence of an accidental approximate global symmetry of the SM known as \emph{custodial SU(2) symmetry}.  The best way to see why the custodial symmetry is important is to consider a beyond-the-SM situation in which it is broken.

\subsection{Scalar triplets and custodial symmetry violation}

Let's consider an extension of the SM Higgs sector containing a triplet of SU(2)$_L$.  A scalar triplet cannot give mass to charged fermions because one can't build an appropriate SU(2)$_L \times$U(1)$_Y$-invariant operator with dimension four; however, a scalar triplet with nonzero vev does contribute to the $W$ and $Z$ boson masses.

To give a scalar triplet a vev without breaking electromagnetism, we have to put the vev in an electrically-neutral component.  There are two choices for the hypercharge assignment of an SU(2)$_L$ triplet that give us a neutral component:
\begin{itemize}
\item $Y=0$, real scalar:
\begin{equation}
	\Xi = \left( \begin{array}{c} \xi^+ \\ \xi^0 \\ \xi^- \end{array} \right)
	\rightarrow \left( \begin{array}{c} \xi^+ \\ \xi^0 + v_{\xi} \\ \xi^- \end{array} \right).
\end{equation}
\item $Y=1$, complex scalar:\footnote{Taking $Y = -1$ would just give us the multiplet that is conjugate to the one with $Y=1$.}
\begin{equation}
	X = \left( \begin{array}{c} \chi^{++} \\ \chi^+ \\ \chi^0 \end{array} \right)
	\rightarrow \left( \begin{array}{c} \chi^{++} \\ \chi^+ \\ v_{\chi} + (h_{\chi} + i a_{\chi})/\sqrt{2} 
	\end{array} \right).
\end{equation}
\end{itemize}

Now consider the terms in the covariant derivative that contribute to the $W$ and $Z$ masses.  The gauge-kinetic Lagrangian for our SM Higgs doublet and the two types of triplets reads
\begin{equation}
	\mathcal{L} \supset ( \mathcal{D}_{\mu} \Phi)^{\dagger} (\mathcal{D}^{\mu} \Phi)
	+ ( \mathcal{D}_{\mu} X)^{\dagger} (\mathcal{D}^{\mu} X)
	+ \frac{1}{2} ( \mathcal{D}_{\mu} \Xi)^{\dagger} (\mathcal{D}^{\mu} \Xi),
\end{equation}
where the covariant derivative is given as usual by
\begin{equation}
	\mathcal{D}_{\mu} = \partial_{\mu} - i g^{\prime} B_{\mu} Y - i g W_{\mu}^a T^a
	= \partial_{\mu} - i g^{\prime} B_{\mu} Y 
	- i g \left[ \frac{1}{\sqrt{2}} (W^+_{\mu} T^+ + W^-_{\mu} T^-) + W^3_{\mu} T^3 \right].
\end{equation}
Consider a \emph{generic} term $(\mathcal{D}_{\mu} X)^{\dagger} (\mathcal{D}^{\mu} X)$ in the gauge-kinetic Lagrangian.  Its contribution to the $W$ and $Z$ masses (i.e., the terms proportional to $v_X^2$) looks like
\begin{eqnarray}
	(\mathcal{D}_{\mu} X)^{\dagger} (\mathcal{D}^{\mu} X) &\supset& 
	X^{\dagger} \left[ \frac{g^2}{2} W^+_{\mu} W^{- \mu} (T^+ T^- + T^- T^+) 
	+ g^2 W^3_{\mu} W^{3 \mu} (T^3)^2 \right. \nonumber \\
	&& \left. \ \ \textcolor{white}{\frac{g^2}{4}}
	+ g^{\prime 2} B_{\mu} B^{\mu} (Y)^2
	+ 2 g g^{\prime} B_{\mu} W^{3 \mu} (Y T^3) \right] X,
\end{eqnarray}
plus terms that will not give $v \cdots v$.

The term $(T^+ T^- + T^- T^+)$ can be evaluated using a trick from quantum mechanics [remember, spin is just SU(2)!]:
\begin{eqnarray}
	T^+ T^- + T^- T^+ &=&
	(T^1 + i T^2) (T^1 - i T^2) + (T^1 - i T^2) (T^1 + i T^2)
	\nonumber \\
	&=& 2 \left[ (T^1)^2 + (T^2)^2 \right] \nonumber \\
	&=& 2 \left[ |\vec T|^2 - (T^3)^2 \right] \nonumber \\
	&=& 2 \left[ T (T+1) - (T^3)^2 \right],
\end{eqnarray}
where in the last step we used $|\vec T|^2 X = T (T+1) X$, where $T$ is the total isospin quantum number of $X$ (1/2 for a doublet, 1 for a triplet).  We'll also use $Q = T^3 + Y$ so that $T^3 = Q - Y = -Y$ for the neutral component of $X$ where the vev lives.

So for the $v \cdots v$ terms only, we have
\begin{eqnarray}
	(\mathcal{D}_{\mu} X)^{\dagger} (\mathcal{D}^{\mu} X) &\supset& 
	X^{\dagger} \left\{ g^2 W^+_{\mu} W^{- \mu} \left[ T(T+1) - Y^2 \right]
	+ g^2 W^3_{\mu} W^{3 \mu} (Y)^2 \right. \nonumber \\
	&& \left. \qquad
	+ g^{\prime 2} B_{\mu} B^{\mu} (Y)^2
	- 2 g g^{\prime} B_{\mu} W^{3 \mu} (Y)^2 \right\} X.
\end{eqnarray}
The pieces are now easy to evaluate for any choice of SU(2)$_L$ and U(1)$_Y$ quantum numbers: 
\begin{center}
\begin{tabular}{lll}
Doublet, $Y = 1/2$: \ \ & $T(T+1) - Y^2 = \frac{1}{2}$, \ \ & $Y^2 = \frac{1}{4}$. \\
Triplet, $Y = 0$: & $T(T+1) - Y^2 = 2$, & $Y^2 = 0$. \\
Triplet, $Y = 1$: & $T(T+1) - Y^2 = 1$, & $Y^2 = 1$. 
\end{tabular}
\end{center}

Now we can work out the contributions of each scalar to the $W$ and $Z$ masses.  For the SM Higgs doublet with $Y = 1/2$, 
\begin{equation}
	\langle \Phi \rangle = \left( \begin{array}{c} 0 \\ v_{\phi}/\sqrt{2} \end{array} \right),
\end{equation}
so that
\begin{equation}
	(\mathcal{D}_{\mu} \Phi)^{\dagger} (\mathcal{D}^{\mu} \Phi) \supset
	\frac{g^2 v_{\phi}^2}{4} W^+_{\mu} W^{- \mu}
	+ \frac{g^2 v_{\phi}^2}{8} W^3_{\mu} W^{3 \mu}
	+ \frac{g^{\prime 2} v_{\phi}^2}{8} B_{\mu} B^{\mu}
	- 2 \frac{g g^{\prime} v_{\phi}^2}{8} B_{\mu} W^{3 \mu}.
\end{equation}
We can write the doublet's contribution to the gauge boson masses-squared in \emph{matrix form} in the basis $(W^1, W^2, W^3, B)$:
\begin{equation}
	M^2_{\Phi} = \frac{v_{\phi}^2}{4} \left( \begin{array}{cccc}
	g^2 & 0 & 0 & 0 \\
	0 & g^2 & 0 & 0 \\
	0 & 0 & g^2 & -g g^{\prime} \\
	0 & 0 & -g g^{\prime} & g^{\prime 2} 
	\end{array} \right).
\end{equation}
There are a couple of things to notice:
\begin{itemize}
\item The lower $2 \times 2$ block is diagonalized by the weak mixing angle, 
\begin{equation}
	\sin\theta_W = \frac{g^{\prime}}{\sqrt{g^2 + g^{\prime 2}}}.
\end{equation}
This diagonalization gives the $Z$ boson and (massless) photon eigenstates.

\item When $g^{\prime} \to 0$ (equivalent to $\cos\theta_W \to 1$), $M_W = M_Z$ and there is a ``rotation'' symmetry $W^1 \leftrightarrow W^2 \leftrightarrow W^3$.  This symmetry in the limit $g^{\prime} \to 0$ is the actual custodial SU(2) symmetry.  It is an accident in the SM (as we'll see when we evaluate the same $W$ and $Z$ boson mass-squared matrix for the triplets), and is only an approximate symmetry because it is broken by the gauging of hypercharge.

\end{itemize}

Now let's consider the real triplet with $Y = 0$.  Its vev is
\begin{equation}
	\langle \Xi \rangle = \left( \begin{array}{c} 0 \\ v_{\xi} \\ 0 \end{array} \right),
\end{equation}
so that
\begin{equation}
	\frac{1}{2} (\mathcal{D}_{\mu} \Xi )^{\dagger} ( \mathcal{D}^{\mu} \Xi) \supset
	g^2 v_{\xi}^2 W^+_{\mu} W^{- \mu}.
\end{equation}
Because $Y = 0$, the coefficients of the $W^3_{\mu} W^{3 \mu}$, $B_{\mu} B^{\mu}$, and $B_{\mu} W^{3 \mu}$ terms are zero!  The $Y=0$ triplet's contribution to the gauge boson masses-squared in matrix form is given by
\begin{equation}
	M^2_{\Xi} = v_{\xi}^2 \left( \begin{array}{cccc}
	g^2 & 0 & 0 & 0 \\
	0 & g^2 & 0 & 0 \\
	0 & 0 & 0 & 0 \\
	0 & 0 & 0 & 0 \end{array} \right).
\end{equation}
In particular, the $Y = 0$ triplet contributes to the $W^{\pm}$ mass but does not contribute to the $Z$ or photon masses.  Notice also that rotating the lower $2 \times 2$ block by the weak mixing angle $\theta_W$ has no effect on this matrix.  This means that when we add up all the contributions to the $W$ and $Z$ masses-squared in matrix form, the contribution from $M^2_{\Xi}$ will not change the value of the mixing angle $\theta_W$ between the $Z$ and the photon: this mixing angle is in fact set entirely by the values of the gauge couplings $g$ and $g^{\prime}$.

However, notice that in the $g^{\prime} \to 0$ limit, the $Y=0$ triplet does \emph{not} generate the same masses for $W$ and $Z$.  There is a symmetry $W^1 \leftrightarrow W^2$, but no longer is there the full custodial SU(2) symmetry $W^1 \leftrightarrow W^2 \leftrightarrow W^3$.

Finally let's consider the complex triplet with $Y = 1$.  Its vev is
\begin{equation}
	\langle X \rangle = \left( \begin{array}{c} 0 \\ 0 \\ v_{\chi} \end{array} \right),
\end{equation}
so that
\begin{equation}
	( \mathcal{D}_{\mu} X)^{\dagger} (\mathcal{D}^{\mu} X) \supset
	g^2 v_{\chi}^2 W^+_{\mu} W^{- \mu}
	+ g^2 v_{\chi}^2 W^3_{\mu} W^{3 \mu}
	+ g^{\prime 2} v_{\chi}^2 B_{\mu} B^{\mu}
	- 2 g g^{\prime} v_{\chi}^2 B_{\mu} W^{3 \mu}.
\end{equation}
The $Y=1$ triplet's contribution to the gauge boson masses-squared in matrix form is given by
\begin{equation}
	M_X^2 = v_{\chi}^2 \left( \begin{array}{cccc}
	g^2 & 0 & 0 & 0 \\
	0 & g^2 & 0 & 0 \\
	0 & 0 & 2 g^2 & -2 g g^{\prime} \\
	0 & 0 & -2 g g^{\prime} & 2 g^{\prime 2} \end{array} \right).
\end{equation}
Notice that the lower $2 \times 2$ block of this matrix is still diagonalized by the same weak mixing angle $\theta_W$ as for the doublet.  The photon is also still massless.  However, as for the $Y=0$ triplet, the $Y=1$ triplet does not generate the same masses for the $W$ and $Z$ in the limit $g^{\prime} \to 0$.  Again there is no custodial SU(2) symmetry $W^1 \leftrightarrow W^2 \leftrightarrow W^3$ in that limit.

The lack of custodial SU(2) symmetry has a profound low-energy experimental consequence: it changes the relative strength of the charged and neutral weak currents.  In the presence of the $Y=0$ and $Y=1$ triplets, we have
\begin{equation}
	M_W^2 = \frac{g^2}{4} (v_{\phi}^2 + 4 v_{\xi}^2 + 4 v_{\chi}^2), \qquad \qquad
	M_Z^2 = \frac{g^2 + g^{\prime 2}}{4} (v_{\phi}^2 + 8 v_{\chi}^2) 
	= \frac{g^2}{4 c_W^2} (v_{\phi}^2 + 8 v_{\chi}^2),
\end{equation}
so that
\begin{equation}
	\rho \equiv \frac{M_W^2}{c_W^2 M_Z^2} 
	= \frac{v_{\phi}^2 + 4 v_{\xi}^2 + 4 v_{\chi}^2}{v_{\phi}^2 + 8 v_{\chi}^2}.
\end{equation}
The rho parameter is measured to be very close to one.  This implies that, if one or both of the triplets that we've discussed in this section are present, their vevs must either be very small compared to $v_{\phi}$, or they must be \emph{tuned} so that $v_{\xi} = v_{\chi} \equiv v_3$.  Tuning the triplet vevs to be equal in this way leads to a contribution to the gauge boson mass-squared matrix from the two triplets of
\begin{equation}
	M^2_{X + \Xi} = 2 v_3^2 \left( \begin{array}{cccc}
	g^2 & 0 & 0 & 0 \\
	0 & g^2 & 0 & 0 \\
	0 & 0 & g^2 & -g g^{\prime} \\
	0 & 0 & -g g^{\prime} & g^{\prime 2} 
	\end{array} \right).
\end{equation}
This matrix has the same form as that for the doublet!  The custodial SU(2) symmetry is restored: in the limit $g^{\prime} \to 0$, the $W$ and $Z$ masses again become equal and we recover the rotation symmetry $W^1 \leftrightarrow W^2 \leftrightarrow W^3$.

\subsection{Restoring the custodial symmetry}

The custodial symmetry can be better understood by studying the global symmetries of the SM Higgs sector.  We can think of the SM Higgs doublet as an object with four real components, as in Eq.~(\ref{eq:phi4component})---i.e., as a four-component vector.  The Higgs potential by itself (not including the gauge interactions) preserves a global O(4) symmetry, which is broken down to O(3) when one component of $\Phi$ gets a vev.  This is a larger symmetry than the gauged SU(2)$_L \times$U(1)$_Y$; in fact, the global O(4) corresponds to a global SU(2)$_L \times$SU(2)$_R$ symmetry, as can be seen by writing $\Phi$ in the form of a \emph{bidoublet}:
\begin{equation}
	\Phi = \left( \begin{array}{cc} \phi^{0*} & \phi^+ \\ -\phi^{+*} & \phi^0 \end{array} \right),
\end{equation}
which transforms under the global SU(2)$_L \times$SU(2)$_R$ symmetry as
\begin{equation}
	\Phi \rightarrow \exp\left( i \alpha_L^a \frac{\sigma^a}{2} \right) \Phi
	\exp \left( -i \alpha_R^a \frac{\sigma^a}{2} \right),
\end{equation}
where $\sigma^a$ are the Pauli matrices as usual and $\alpha_L^a$ and $\alpha_R^a$ are constants.  Promoting this global symmetry to the local (gauge) symmetry of the SM, all three generators $\alpha_L^a$ of the SU(2)$_L$ global symmetry become spacetime-dependent functions $\lambda_L^a(x)$, but only the third generator $\alpha_R^3$ of the SU(2)$_R$ global symmetry is gauged, becoming the $\lambda_Y(x)$ generator of hypercharge.  Gauging only one component of a global symmetry violates the global symmetry.

In any case, when $\Phi$ gets a vev, the bidoublet becomes
\begin{equation}
	\langle \Phi \rangle = \frac{1}{\sqrt{2}} \left( \begin{array}{cc} v_{\phi} & 0 \\ 0 & v_{\phi}
	\end{array} \right),
\end{equation}
which is proportional to the unit matrix.  This breaks the global SU(2)$_L \times$SU(2)$_R$ symmetry down to the \emph{diagonal subgroup} SU(2)$_{\rm diag}$.  This diagonal subgroup is the custodial SU(2) symmetry.  It is an accidental (just a consequence of the ability to write $\Phi$ as a bidoublet and have its vev automatically proportional to the unit matrix), approximate (violated by hypercharge gauge interactions, as well as by the fact that the Yukawa couplings of the up-type and down-type fermions are different) global symmetry of the SM.

We can take advantage of this observation to \emph{engineer} the relationship $v_{\xi} = v_{\chi}$ between the triplet vevs by putting together the complex $Y=1$ triplet and the real $Y=0$ triplet into a $3\times 3$ object that transforms as a triplet under both the global SU(2)$_L$ and SU(2)$_R$ symmetries:
\begin{equation}
	\tilde X = \left( \begin{array}{ccc} \chi^{0*} & \xi^+ & \chi^{++} \\
	\chi^- & \xi^0 & \chi^+ \\
	\chi^{--} & \xi^- & \chi^0 \end{array} \right),
\end{equation}
where $\chi^- = - \chi^{+*}$ and $\xi^- = -\xi^{+*}$.  If the scalar potential is constructed to preserve the global SU(2)$_L \times$SU(2)$_R$ symmetry (this can be done by eliminating some terms in the most general gauge invariant scalar potential for two triplets and a doublet; the resulting model is called the Georgi-Machacek model~\cite{GM}), then it is natural for the triplets to get a vev of the form
\begin{equation}
	\langle \tilde X \rangle = \left( \begin{array}{ccc} v_{\chi} & 0 & 0 \\
	0 & v_{\chi} & 0 \\
	0 & 0 & v_{\chi} \end{array} \right),
\end{equation}
again proportional to the unit matrix.  This preserves the same diagonal subgroup SU(2)$_{\rm diag}$ that is automatically preserved when the SM Higgs doublet gets a vev.  Again, the symmetry is only approximate: one-loop corrections involving hypercharge gauge interactions regenerate the SU(2)$_R$-violating terms that were left out of the scalar potential (but this is not so bad if the cutoff scale is not too far above the electroweak scale).

In this way, scalar triplets with a sizable vev can be added to the SM, allowing for some interesting beyond-the-SM Higgs phenomenology.

\section{Outlook}
\label{sec:summary}

It has become a clich\'e to say that the Higgs boson discovery opens a new era in experimental particle physics.  That is because it is true.  Detailed studies of the Higgs boson's properties will dominate the LHC program over the next ten years, and form an important part of the physics case for the high-luminosity LHC upgrade to run in the next ten years after that.  High-precision Higgs studies are the primary motivation for the construction of the International Linear Collider.  This experimental program will allow the Higgs couplings to be measured with a precision reaching the sub-percent level, providing sensitivity to new physics affecting the Higgs sector up to energy scales as high as a few TeV.

But the Higgs is also a genuinely new beast from a theoretical point of view.  If it is truly a fundamental\footnote{The alternative to a fundamental scalar is that the Higgs is a composite (bound state) of some new fundamental fermions, confined by a new gauge interaction.  This scenario is sometimes called Technicolor.} scalar particle, it is the first one ever observed in nature.  Higgs boson decays to fermions, and Higgs production via gluon fusion, represent the first observation of a collection of brand new interactions (the Yukawa couplings).  The fact that the Higgs is condensed in the vacuum---i.e., that its vacuum expectation value is nonzero---implies the existence of a brand new quartic scalar interaction (the Higgs self-coupling).  It is not an exaggeration to say that these are the first interactions ever observed by humans that are not a consequence of gravity or the gauge forces.

The great triumphs of the Standard Model---QED, precision electroweak measurements, perturbative and nonperturbative (lattice) QCD, and the CKM framework for flavor physics---are all consequences of the gauge principle.  This we understand.  The great mysteries of the Standard Model---the origin of the $W$ and $Z$ boson masses, the origin of the quark and lepton masses along with their mixing and CP violation, the origin of neutrino masses and their mixing, dark energy and inflation, and the hierarchy problems of the electroweak-breaking scale and the cosmological constant---all have something to do with what could be called the properties of the vacuum.  The Higgs is our first tangible piece of the vacuum, in that it is a quantum of the field that is condensed in the vacuum.  It could teach us some of nature's deepest secrets.

\section{Homework questions}
\label{sec:homework}

\begin{enumerate}

\item
Compute the tree-level decay partial width for the Higgs boson into a
pair of bottom quarks and show that it is given by
\begin{equation}
  \Gamma(h \to b \bar b) = \frac{N_c}{8 \pi} \frac{m_b^2}{v^2} m_h
  \left[ 1 - \frac{4 m_b^2}{m_h^2} \right]^{3/2},
\end{equation}
where $N_c = 3$ is the number of colours of the $b$ quark 
and $v = 246$~GeV is the Higgs vacuum expectation value (vev).
The Feynman rule for the $h b \bar b$ vertex is $-i m_b/v$.

\item
Imagine that the scalar potential for the Standard Model Higgs field 
contained a $\phi^6$ term, as follows:
\begin{equation}
  V(\Phi) = - \mu^2 \Phi^{\dagger} \Phi
  + \lambda (\Phi^{\dagger} \Phi)^2
  + \frac{1}{\Lambda^2} (\Phi^{\dagger} \Phi)^3,
\end{equation}
where $\mu^2$, $\lambda$, and $1/\Lambda^2$ are all positive and $\Lambda$
has dimensions of mass.

Minimize the potential and eliminate $\mu^2$ in favor of the Higgs
vev $v$.  Then find the Higgs mass and the Feynman rules for the $hhh$
and $hhhh$ coupling vertices in terms of $v$, $\lambda$, and $1/\Lambda^2$.
You can work in the unitarity gauge and write 
\begin{equation}
  \Phi = \left( \begin{array}{c} 0 \\ (h + v)/\sqrt{2} \end{array} \right).
\end{equation}
(Recall that when the $1/\Lambda^2$ term is not there, the Higgs mass is $m_h = \sqrt{2 \lambda} v$, the $hhh$ coupling Feynman rule is $-3im_h^2/v$, and the $hhhh$ coupling Feynman rule is $-3im_h^2/v^2$.  The idea here is to see whether you can tell that the $1/\Lambda^2$ term is there by measuring $m_h$ and the $hhh$ coupling and comparing to the Standard Model relationship.)

\item
The LHC measures rates for Higgs boson production and decay into specific final states, which can be written in the ``zero width approximation'' as
\begin{equation}
	{\rm Rate}_{ij} = \sigma_i \times {\rm BR}_j 
	= \sigma_i \times \frac{\Gamma_j}{\Gamma_{\rm tot}}.
\end{equation}
If there is a new, non-SM decay mode of the Higgs, which is unobservable at the LHC (for example, Higgs decay into light-quark jets, which would be buried under background), all the observable Higgs signal rates can be kept the same by cranking up the production couplings at the same time as the branching ratio to the new final state is increased.  All couplings must be increased by the same factor in order to keep the ratios of rates fixed; we can denote that factor by $\kappa$, in which case 
$\sigma_i = \kappa^2 \sigma_i^{\rm SM}$ and $\Gamma_j = \kappa^2 \Gamma_j^{\rm SM}$.

Work out the relationship between the coupling scaling factor $\kappa$ and the new unobservable decay branching ratio BR$_{\rm new}$ that is required to keep all the Higgs signal rates fixed to their SM values.  (This relation defines a ``flat direction'' in the Higgs coupling fit using LHC data.  To cut off the flat direction, fits to LHC Higgs data usually assume either that there are no unobservable new decay modes (i.e., BR$_{\rm new} = 0$) or that the Higgs couplings to $WW$ and $ZZ$ cannot be larger than their SM values (i.e., $\kappa \leq 1$ in our notation).  The latter happens to be true in all models containing only Higgs doublets and/or singlets of SU(2)$_L$.)

\item
The correct ratio for the $W$ and $Z$ masses can (by coincidence) also be generated by a septet (or seven-plet) of SU(2)$_L$ with hypercharge $Y = 2$ (in the convention $Q = T^3 + Y$).  If the vev of the septet is given by (note that $v_7$ is in the neutral component, $T^3 = -2$, so that $Q = T^3 + Y = 0$)
\begin{equation}
	\langle \chi_7 \rangle = \left( \begin{array}{c} 0 \\ 0 \\ 0 \\ 0 \\ 0 \\ v_7 \\ 0 \end{array} \right),
\end{equation} 
work out the gauge boson mass-squared matrix in the $W^1, W^2, W^3, B$ basis and show that it is indeed proportional to the SM case,
\begin{equation}
	M^2_{\rm SM} = \frac{v_{\rm SM}^2}{4} \left( \begin{array}{cccc}
	g^2 & 0 & 0 & 0 \\
	0 & g^2 & 0 & 0 \\
	0 & 0 & g^2 & -g g^{\prime} \\
	0 & 0 & -g g^{\prime} & g^{\prime 2} \end{array} \right).
\end{equation}
If the $W$ and $Z$ masses were generated entirely by the septet, what value of $v_7$ would be needed?  (Recall that $M_W = g v_{\rm SM}/2$ and $M_Z = \sqrt{g^2 + g^{\prime 2}} v_{\rm SM}/2$ at tree level for $v_{\rm SM} = 246$~GeV.)

Could all mass generation in the Standard Model be accomplished by the septet?  Why or why not?

\end{enumerate}

\section*{Acknowledgments}
I thank the TASI 2013 program co-directors, Iain Stewart and Bogdan Dobrescu, for giving me the opportunity to present these lectures, and the local organizers Tom DeGrand, K.T.~Mahanthappa, and Susan Spika for the smooth running of the summer institute.  I also thank the organizers of the Tri-Institute Summer School on Elementary Particles 2013 at TRIUMF in Vancouver, Canada, for which the homework questions were developed.
TASI was supported by the U.S.~Department of Energy and National Science Foundation.
This work was also supported by the Natural Sciences and Engineering Research 
Council of Canada.


\end{document}